\documentclass[a4paper,11pt]{article}
\pdfoutput=1 

\usepackage{jcappub} 
                                        
\usepackage{aas_macros}
\usepackage{subcaption}
\usepackage[T1]{fontenc} 
\usepackage{float}
\usepackage[normalem]{ulem}

\newcommand{\avg}[1]{\ensuremath{\left\langle \,#1\, \right\rangle}}

\newcommand{\Mh}{\ensuremath{h^{-1}M_{\odot}}}

\newcommand{\Mpch}{\ensuremath{h^{-1}{\rm Mpc}} }
\newcommand{\hMpc}{\ensuremath{h\,{\rm Mpc}^{-1}}}

\newcommand{\kms}{\ensuremath{{\rm km\,s}^{-1}}}
\newcommand{\fcoll}{\ensuremath{f_{\text{coll}}}}

\newcommand{\fcollm}{\ensuremath{f_{\text{coll}}^M} }

\newcommand{\be}{\begin{equation}}
\newcommand{\ee}{\end{equation}}

\title{\boldmath Accelerating HI density predictions during the Epoch of Reionization using a GPR-based emulator on N-body simulations}

\usepackage{natbib}
\author[a, 1]{Gaurav Pundir,\note{Corresponding author.}}
\author[b]{Aseem Paranjape,}
\author[c]{and Tirthankar Roy Choudhury}

\affiliation[a]{Department of Physics, \\ 
Indian Institute of Science Education and Research Pune, \\
Dr. Homi Bhabha Road, Pashan, Pune 411008, India}
\affiliation[b]{
Inter-University Centre for Astronomy \& Astrophysics, \\
Post Bag 4, Ganeshkhind, Pune 411007, India}
\affiliation[c]{
National Centre for Radio Astrophysics, TIFR, \\
Post Bag 3, Ganeshkhind, Pune 411007, India}

\emailAdd{gaurav.pundir@students.iiserpune.ac.in}
\emailAdd{aseem@iucaa.in}
\emailAdd{tirth@ncra.tifr.res.in}

\abstract{Building fast and accurate ways to model the distribution of neutral hydrogen during the Epoch of Reionization (EoR) is essential for interpreting upcoming 21 cm observations. A key component of semi-numerical models of reionization is the collapse fraction field $\fcoll(\mathbf{x})$, which represents the fraction of mass within dark matter halos at each location. Using high-dynamic range N-body simulations to obtain this is computationally prohibitive and semi-analytical approaches, while being fast, end up compromising on accuracy. In this work, we bridge the gap by developing a machine learning model that can generate \fcoll\ maps by sampling from the full distribution of \fcoll\ conditioned on the dark matter density contrast $\delta$. The conditional distribution functions and the input density field to the model are taken from low-dynamic range N-body simulations that are more efficient to run. We evaluate the performance of our ML model by comparing its predictions to a high-dynamic range N-body simulation. Using these \fcoll\ maps, we compute the HI and HII maps through a semi-numerical code for reionization. We are able to recover the large-scale HI density field power spectra $(k \lesssim 1\ \hMpc)$ at the $\lesssim10$\% level, while the HII density field is reproduced with errors well below 10\% across all scales. Compared to existing semi-analytical prescriptions, our approach offers significantly improved accuracy in generating the collapse fraction field, providing a robust and efficient alternative for modeling reionization.}

\begin{document}
\maketitle
\flushbottom

\section{Introduction}
\label{sec:intro}

The Epoch of Reionization (EoR) marks an important period in the history of the universe when the first luminous objects ionized the neutral hydrogen (HI) in the intergalactic medium (IGM). Studying this era is crucial for understanding many astrophysical processes, including the emergence of the first stars and galaxies and the growth of cosmic structure (for recent reviews, see \cite{Gnedin22_review, TRC22_review}).
The observational signatures of EoR are extremely faint because of the large distances involved and are also buried under much stronger astrophysical foregrounds. One of the most promising probes is the 21 cm brightness temperature fluctuation, which is a tracer of the HI density fluctuations during EoR \cite{furlanetto06_review,  pritchard12_review, mesinger19_review}.  This has been targeted by radio interferometers such as GMRT\footnote{\href{https://www.gmrt.ncra.tifr.res.in/}{\texttt{https://www.gmrt.ncra.tifr.res.in/}}}, MWA\footnote{\href{https://www.mwatelescope.org/}{\texttt{https://www.mwatelescope.org/}}}, PAPER\footnote{\href{http://eor.berkeley.edu/}{\texttt{http://eor.berkeley.edu/}}}, LOFAR\footnote{\href{http://www.lofar.org/}{\texttt{http://www.lofar.org/}}} and will also be observed by the upcoming HERA\footnote{\href{https://reionization.org/}{\texttt{https://reionization.org/}}} Phase-II and SKA\footnote{\href{https://www.skao.int/en}{\texttt{https://www.skao.int/en}}}. 

In standard models of the EoR that assume galaxies to be the dominant contributors of ionizing photons, reionization proceeds via the formation of `ionized bubbles' containing ionized hydrogen (HII). By modeling the distribution of these ionized bubbles, we can get the distribution of neutral hydrogen, which in turn provides information regarding fluctuations in the 21 cm signal. The most accurate way to achieve this is to run radiative transfer (RT) simulations that take into account the detailed physical interactions between matter and the photons emitted by the sources \cite{gnedin00, ciardi_FW03, iliev06, trac&cen07, petkova&springel09, gnedin14, pawlik17, rosdahl18, kannan21, lewis22}. However, these simulations must have a sufficiently large volume to achieve statistical convergence on the bubble distribution at large scales \cite{iliev14_size, kaur20_size}. Simultaneously, they need to resolve the smallest mass halos capable of forming the first galaxies (typically down to $\sim 10^8\ \Mh$) due to their significant contribution to the ionizing photon budget. This `high-dynamic range' requirement adds significantly to their computational cost and makes them highly inefficient to explore the parameter space of EoR models.

One gets around this problem by resorting to the much faster but approximate semi-numerical models of reionization. These aim to predict the `ionization field' -- describing the fraction of hydrogen ionized at each location -- by using the excursion-set approach \cite{BCEK_91}  and a simple photon counting argument to define the barrier \cite{FZH04}, thus bypassing the complicated radiative transfer physics \cite{Mesinger_2007, zahn07, choudhury09_ES, 21cmfast, lin16_ES, script}. These semi-numerical methods provide a reasonable match to RT simulations in terms of various statistics such as the neutral fraction, bubble size distribution, power spectrum of the ionization field, and so on \cite{majumdar_14, zahn_11}. When used along with semi-numerical galaxy formation codes, the input to these models can be the stellar mass \cite{mutch_16} or the number of ionizing photons entering the IGM at each cell \cite{kim_13}, with the outpt being the ionization fraction in the cell. When used in conjunction with dark-matter-only simulations, the required input is the `collapse fraction field' denoted by $\fcoll(\mathbf{x})$, which is equal to the fraction of dark matter mass within halos in the grid cell at $\mathbf{x}$. This can be computed by first filtering individual halos using the excursion-set formalism from a dark matter density field evolved using Lagrangian perturbation theory \cite{Mesinger_2007, santos_10}. Alternatively, it can be prescribed semi-analytically without explicitly identifying sub-grid halos from the conditional Press-Schechter (hereafter conditional PS) halo mass function \cite{PS_74, BCEK_91}, conditioned on the dark matter density contrast $\delta (\mathbf{x})$ for each cell. One can also use the conditional Sheth-Tormen (hereafter conditional ST) mass function, which is based on the more general ellipsoidal collapse model \cite{ST_99, ST_02}. 

However, these analytical mass functions do not capture the full complexity of halo formation, are not universal and are only an approximate match to N-body simulation results \cite{reed07, tinker08, courtin10, crocce11, bhattacharya11}. In particular at the redshifts of our interest, the conditional PS mass function underestimates the abundance of high mass halos $(M \gtrsim 10^{10} ~\Mh \text{ at } z=7)$ and overpredicts the abundance at low masses $(M \lesssim 10^8 ~\Mh)$. While the ST mass function provides a better match, it still overpredicts the number of very massive halos at high redshifts \cite{reed07}. These considerations are important in studies of reionization since correctly predicting the abundance of halos is crucial for obtaining accurate ionized regions. Therefore, as the first step, one should transition away from the conditional PS and ST mass functions and use N-body simulations to calculate the conditional mass function empirically. However, these approaches only assign the mean \fcoll\ conditioned on the density value of each cell $\avg{\fcoll | \delta}$, whereas in reality, the \fcoll\ value can stochastically fluctuate across different cells with the same density value. Ignoring this `scatter' or `stochasticity' in the collapse fraction (which is primarily due to a dependence of \fcoll\ on environmental variables other than the grid-scale $\delta$) can lead to inaccurate recovery of the small-scale features in the HI maps, as we show later in the paper.  Hence, as the next step, one should use the conditional cumulative distribution function of \fcoll\ conditioned on the density contrast, $\text{CDF}(\fcoll | \delta)$ to sample the \fcoll\ field.

In either case, it is still important for the N-body simulations to have a high-dynamic range. This makes them computationally very expensive and thus one must explore alternatives to enable fast predictions of collapse fraction and subsequently the HI density fields. Attempts to resolve this issue have involved running low-resolution, large-volume simulations and using a high-resolution, small-volume simulation to populate the otherwise unresolved halos. This has been implemented in \cite{ahn12, iliev14_size}, although while not taking into account the scatter in the halo numbers for a given overdensity. Poisson fluctuations in the halo number count around the mean value predicted by the analytical conditional mass functions have been incorporated in certain studies \cite{mcquinn07, santos_10, doussot22}, but this has the limitation of only being valid for large enough cell sizes \cite{sheth_lemson_poisson99A, sheth_lemson_poisson99B}. An alternative approach is to identify matching cells in the small-volume, high-resolution simulation and use halos from these cells to populate the low-resolution box \cite{barsode24}. However, this method requires simultaneous access to both the large-volume and small-volume simulations during the construction of the effective high-dynamic-range box.

In this work, we aim to fully incorporate the effects of stochasticity in the collapse fraction values, by directly using the full $\text{CDF}(\fcoll | \delta)$ obtained from an N-body simulation for sampling the \fcoll\ field. We still use a hybrid scheme of combining information from computationally inexpensive low-dynamic range boxes to mimic a high-dynamic range one, but do so using a machine learning algorithm based on Gaussian Process Regression (GPR). For the sake of comparison, we define this to be the \textit{stochastic} case and also define the \textit{deterministic} case in which \fcoll\ predictions are made by simply assigning the conditional means $\avg{\fcoll | \delta}$. We use the \fcoll\ fields from both the cases as inputs to a semi-numerical code for reionization to obtain the HI and HII maps, and compare the results with those obtained from the \fcoll\ field of a high-dynamic range simulation (ground truth). While we obtain the results for both the cases, the main focus of the paper and the machine learning model is the stochastic case. Therefore, this work aims to establish an ML framework for efficiently modeling fields relevant to EoR by bypassing the need to run a high dynamic range N-body simulation, while improving upon the accuracy of semi-analytical prescriptions.

The details of the simulations used are presented in section \ref{sec:sims}, followed by the ML methodology in section \ref{sec:method}. We show the power spectra results for the HI and HII density fields during EoR obtained using both cases and compare them with the high-dynamic range simulation results in section \ref{sec:results}. Additionally, we also compare the performance of the conditional PS and ST mass functions in predicting the \fcoll, HI and HII fields with the ML method. We discuss some features of our ML model in section \ref{sec:disc} and conclude by summarizing the work and addressing the future directions in section \ref{sec:conc}. The appendices provide additional checks on some of the parameter choices made while building the ML model.

\section{Simulations}
\label{sec:sims}

Here, we describe the various N-body simulation boxes that are used for training, sampling, and benchmarking the ML model. All of these were run using the GADGET-2\footnote{\href{https://wwwmpa.mpa-garching.mpg.de/gadget/}{\texttt{https://wwwmpa.mpa-garching.mpg.de/gadget/}}} code \cite{GADGET}, assuming a flat $\Lambda$CDM cosmology with $H_0=67.8$ \kms Mpc$^{-1}$, $\Omega_m = 0.308$, $\Omega_b=0.04$, $\sigma_8=0.829$, $n_s=0.961$. On the simulation snapshots at the redshifts of interest, we compute the dark matter overdensity field $\delta (\mathbf{x})$ over a default grid size of $\Delta x = 0.5$ \Mpch, using a cloud-in-cell mass-assignment scheme. We then run the Friends-of-Friends (FoF) \cite{FoF} halo finder on these snapshots (excluding the LB box) to get the discrete halo field. The collapse fraction field $\fcoll (\mathbf{x})$ is defined as
\begin{equation}\label{fcoll_eqn}
    \fcoll (\mathbf{x}) = \dfrac{\sum_h m_h (\mathbf{x})}{M_{\text{tot}}(\mathbf{x})}\,,
\end{equation}
where the summation runs over the mass of all the halos $m_h (\mathbf{x})$ contained in the cell at $\mathbf{x}$, and $M_{\text{tot}}(\mathbf{x})$ is the total dark matter mass in the same cell. This field is then computed over the same grid as the density field $\delta(\mathbf{x})$. For the default case, we use 10 as the minimum number of particles for identifying a halo, which corresponds to a minimum halo mass of $4.08 \times 10^8\ \Mh$ for both the SB and RB as defined below, since they have the same particle mass resolution. 
\begin{itemize}
    \item Small Boxes (SB): These have a volume of $V=(40\ \Mpch)^3$ and contain $N = 512^3$ particles. 7 realizations of these are run with different seeds, and for each, the overdensity and collapse fraction fields are computed. These pairs of $(\delta, \fcoll)$ found for each cell are then combined over all cells and over all 7 realizations to get a list of $80^3 \times 7 = 3584000$ $(\delta, \fcoll)$ pairs, from which the training data is constructed (refer subsection \ref{subsec:binning}). Each realization of these simulations took $\sim 210$ CPU hours to run, consuming a maximum RAM of around 20 GB. 

    \item Reference Box (RB): This box has a volume of $V=(80\ \Mpch)^3$ and number of particles $N=1024^3$. With both the volume and the number of particles 8 times greater than the SBs, it has the same particle mass resolution $(M_{p\text{, min}})$ as them (since $M_{p\text{, min}} \propto \frac{V}{N}$), and consequently the same minimum halo mass as well. This box is our `ground truth' -- the goal of our emulator will be to recover the statistics of this high dynamic range box. This simulation took $\sim 2900$ CPU hours to run, consuming a maximum RAM of 160 GB.

    \item Large Box (LB): This box has a volume of $V=(80\ \Mpch)^3$ and number of particles $N=512^3$. Therefore, it has a coarser particle resolution than the SBs, but the same volume as the RB. This box is used solely to provide the density values to be input into the emulator and make the \fcoll\  predictions to be compared with the ground truth RB, and hence we do not run a halo finder on it. This simulation took $\sim 220$ CPU hours to run, consuming a maximum RAM of around 20 GB. Note that the combination of SB and LB requires significantly lesser RAM (20 GB) as compared to running the RB (160 GB).

\end{itemize}

\section{Methodology}
\label{sec:method}

From the SB simulation boxes outlined in section \ref{sec:sims}, we obtain the $(\delta, \fcoll)$ pairs for each cell. We can then bin the $\delta$ values from the SBs, and collect the \fcoll\ values falling in each bin to either (a) compute their conditional mean $\avg{\fcoll|\delta}$ or (b) construct the conditional cumulative distribution function $\text{CDF}(\fcoll| \delta)$. Using (a) and (b) to make the \fcoll\ predictions precisely corresponds to the deterministic and stochastic cases as defined at the end of section \ref{sec:intro}, respectively. The GPR training as described in the next subsections is required only for the stochastic case.

\subsection{Binning}
\label{subsec:binning} 

The goal is to use the emulated CDF to directly sample an \fcoll\ value, if a new $\delta$ value is given as the input. This amounts to the assumption that the spatial distribution of collapse fractions is primarily dictated by the local overdensity, and the cumulative effect of other environmental factors is modeled by random sampling from the conditional CDFs.

The binning of the overdensity values is made trickier by their highly skewed distribution since extremely low and high values are quite rare. If a uniform binning scheme is adopted, to accurately capture the variation of the conditional CDF between two intermediate $\delta$ values, the bin width must be made sufficiently small. This causes too few \fcoll\ values to be found in higher $\delta$ bins, leading to a very noisy CDF. Thus, to strike a balance between noise and systematic error, we adopt a variable binning scheme, where the bin width is set to a reference value at $\delta=0$, and it increases along either direction. The bins are defined in $\log(1+\delta)$, and usually have a reference value of around 0.03 dex at $\delta=0$. The other parameter that we must decide in the training data is the number of bins in \fcoll\ used to make the CDFs for a fixed $\delta$ bin. This, along with the $\delta$ bin widths at the two extremes are optimized for each case that we present separately. The optimal extreme bin widths are around $\sim 0.05$ dex and $\sim 0.2$ dex, while the optimal number of \fcoll\ bins is either 500 or 900, depending on the case. We refer the reader to Appendix \ref{appendix:optimization of binning} for more details on the optimization procedure, where we also study the effect of using a fixed binning scheme (optimized for the default $z=7$ case) directly on the other cases. 

\subsection{Training using Gaussian Process Regression}
\label{training}

We employ the Gaussian Process Regression (GPR) technique to construct our interpolator function. This is a non-parametric method that approximates the collection of the target function values $\mathbf{y}$ as a \textit{Gaussian Process} over the inputs $\mathbf{x}$, specified by a mean function, $\mu (\mathbf{x})$ and a covariance function, $k(\mathbf{x}, \mathbf{x'})$. This is a standard regression algorithm described in Rasmussen \& Williams \cite{Rasmussen2006Gaussian} and has been implemented in Scikit-Learn\footnote{\href{https://scikit-learn.org/}{\texttt{https://scikit-learn.org/}}}. The mean function is usually taken to be $\mathbf{0}$ in the prior after appropriate normalization of the data. In our case, we use the anisotropic Matérn kernel with $\nu = 2.5$ as the covariance function. Training the GPR model then entails learning the values of the \textit{hyperparameters} associated with the Matérn kernel. We opt for GPR as our interpolation method due to its relatively better ability to generalize to the case of a greater number of conditioning variables beyond $\delta$ alone. This constitutes future work and is addressed at the end of section \ref{sec:conc}.

The hyperparameter optimization is carried out using an anisotropic simulated annealing (ASA) procedure, following \cite{picasa}. As demonstrated in previous studies of reionization \cite{barun, tirth_aseem_barun}, this method works well for GPR hyperparameter training while preventing issues faced by some other Scikit-Learn methods of getting stuck at a local minimum of the cost function where the corresponding GPR is suboptimal in performance. We briefly outline the details of the algorithm for completeness. First, the data is divided into two parts -- one for training and another for validation. Once the training data is specified, the ASA procedure involves evaluating the log marginal likelihood using algorithm 2.1 of \cite{Rasmussen2006Gaussian} over a region with sparsely distributed values in hyperparameter space, which is then iteratively refined to zoom-in on the region of hyper-likelihood maximum (or equivalently, the cost function minimum). 

The hyperparameter vector $\mathbf{h}$ that minimizes the cost function is then used to make predictions on the validation data, again following algorithm 2.1 of \cite{Rasmussen2006Gaussian} as implemented in Scikit-Learn. A convergence criterion is defined by requiring the magnitude of the 1st and 99th percentiles of $\hat{\alpha} - \alpha$ to be less than a threshold set by the user, called \verb|cv_thresh| (here $\hat{\alpha}$ is the predicted value of the function and $\alpha$ is the true value at the same input). If the convergence criterion is not satisfied, the entire process repeats with a training data larger in size by 10\%, and this cycle continues until the maximum number of iterations or $\geq 80$\% of the full data is used for training. \verb|cv_thresh| is usually taken to be around 0.015 in our case. 

Once the training is complete, we have a properly trained interpolator function at our disposal, that can be used to sample the values of the function $\hat{\alpha}$ at any desired input. We use the GPR training to emulate the $\text{CDF}(\fcoll| \delta)$ as obtained in the previous subsection, viewed as a function of \fcoll\ and $\delta$, thereby setting $\alpha = \text{the CDF value}$. For most cases, the training ends within $\sim 10$ minutes on 4 CPU cores and uses around 10-15\% of the full data for training. 

\subsection{Sampling}

Our idea is to be able to recover the \fcoll\ field of RB by a combination of information from SB and LB. We have used SB for obtaining the conditional means and the conditional CDFs to train the GPR, and we now use the $\delta$ values from the LB as the corresponding input to make the \fcoll\ predictions. 

For a given input $\delta_0$ from the LB, we return (a) the conditional mean $\avg{\fcoll | \delta_m}$ for the deterministic case, where $\delta_m$ is the middle value of the bin that contains $\delta_0$ and (b) an \fcoll\ value randomly drawn from the emulated $\widehat{\text{CDF}}(\fcoll | \delta_0)$ for the stochastic case. We use inverse transform sampling for this case where we draw a random number between 0 and 1 and return the smallest \fcoll\ at which $\widehat{\text{CDF}}(\fcoll | \delta_0)$ equals the random number. We can see that the latter method naturally accounts for scatter in the \fcoll\ predictions for a fixed $\delta$ while the former does not. This sampling is done on a cell-by-cell basis, to produce a prediction of \fcoll\ for each cell based on its $\delta$~value in LB.

\section{Results}
\label{sec:results}

In this section, we benchmark the various \fcoll\ predictions against the ground truth taken from the RB. Our primary interest lies in modeling the neutral hydrogen density field during the Epoch of Reionization (EoR), and we obtain this from the collapse fraction field by using \textbf{S}emi-numerical \textbf{C}ode for \textbf{R}e\textbf{I}onization with \textbf{P}ho\textbf{T}on-conservation \textsc{(script)}\footnote{\href{https://bitbucket.org/rctirthankar/script}{\texttt{https://bitbucket.org/rctirthankar/script}}} \cite{script}. The section is divided into two parts - Fiducial, where we discuss the \fcoll\ and \textsc{script} results corresponding to the fiducial choice of parameters and Variation, where we compare the \textsc{script} results for the semi-analytical methods and extend them to variations in the parameters.

\subsection{Fiducial}\label{subsec:fiducial}

\subsubsection{\fcoll\ results}

\begin{figure}[h]
    \centering
    \begin{subfigure}{0.49\linewidth}
    \centering
    \includegraphics[width=\linewidth]{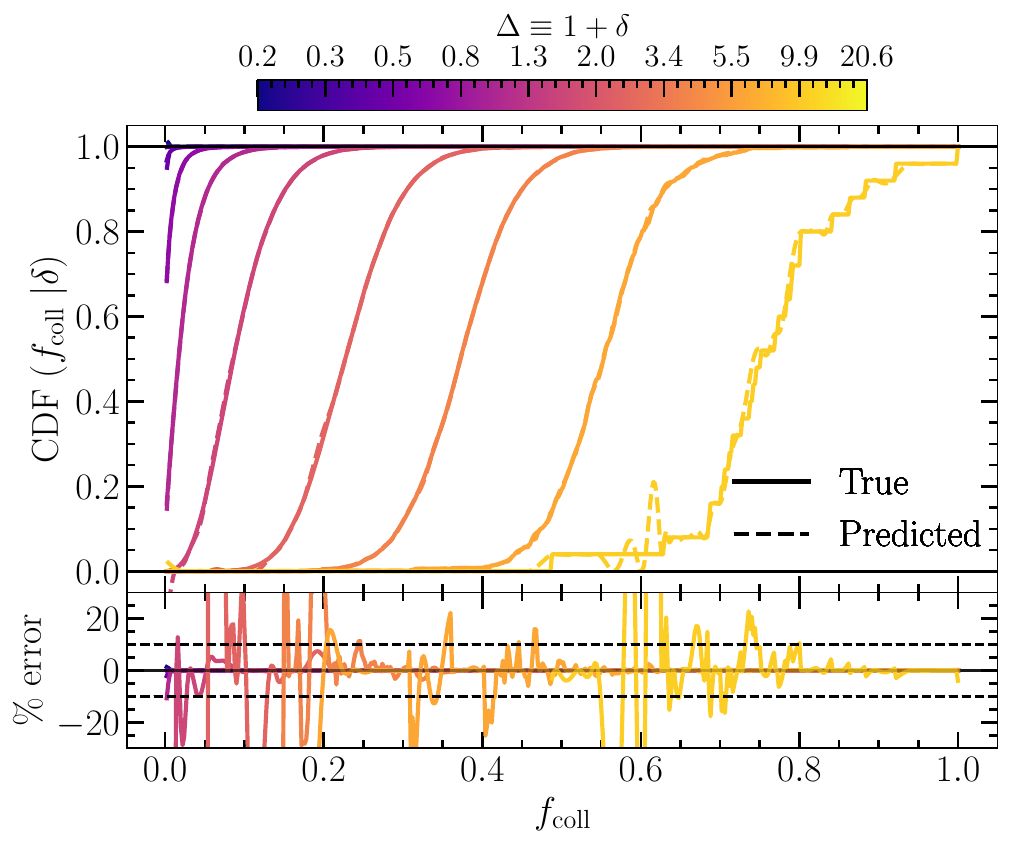}
    \caption{}
    \label{pred_vs_true_CDF}
\end{subfigure}
\hspace{0em}
    \begin{subfigure}{0.49\linewidth}
    \centering
    \includegraphics[width=\linewidth]{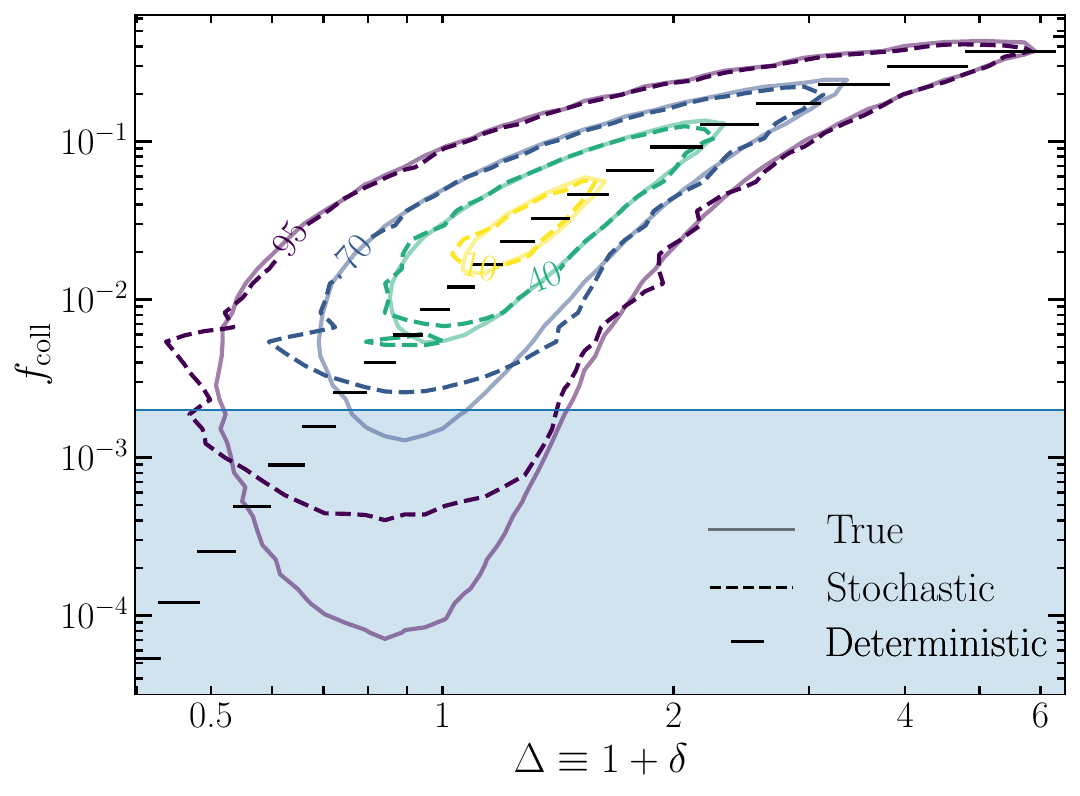}
    \caption{}
    \label{joint_distri}
    \end{subfigure}
    
    \caption{\small(a) Comparison between the true CDF from the training data and the interpolator's prediction, shown at 10 different $\Delta$~values. The relative error occasionally blows up due to the small values of the CDFs, (b) Comparison of the joint distribution of \fcoll\ and $\Delta$. The 10, 40, 70 and 95 percentile contours are shown and the blue region demarcates the \fcoll\ less than the first bin edge defined during the \fcoll\ binning. The conditional means calculated in the deterministic case for each delta bin are also shown using black horizontal lines. This shows that over most of the \fcoll\ and $\Delta$ range, our interpolator recovers the true distribution to a high accuracy.  } 
\end{figure}

We consider our fiducial case to have redshift $z=7$, grid size $\Delta x = 0.5$ \Mpch, and minimum halo mass $M_{h, \text{min}} = 4.08 \times 10^8$ \Mh~(corresponding to 10 particles per halo for the SB and RB).
Henceforth, we shall use the notation $\Delta \equiv 1 + \delta$. We first look at the results of GPR training, by comparing the emulated and training CDFs as a function of \fcoll\ conditioned on 10 different $\Delta$~values, as shown in Figure \ref{pred_vs_true_CDF}. It can be seen that both the training and prediction CDF become noisy at very high $\Delta$, due to a smaller number of \fcoll\ values. 

The recovery of the joint distribution of non-zero \fcoll\ values and their corresponding $\Delta$~is shown in Figure \ref{joint_distri}. While the contours are very similar between truth and prediction at intermediate to high \fcoll\ and $\Delta$, the very low \fcoll\ values are not recovered as well. We understand this to be a limitation of the way we set up the training data for the GPR, where the smallest value of the training CDF that is fed into the GPR is CDF$(\fcoll\ = 0.002)$. The region below $\fcoll\ = 0.002$ is shaded in blue. The interpolator ends up overestimating the CDF at \fcoll\ below this threshold and that leads to an oversampling of $\fcoll\ = 0$ values, and consequently an undersampling of very low $\fcoll\ \lesssim 10^{-3}$. Attempting to fix this problem by incorporating CDF$(\fcoll\ = 0)$ during the training does not provide any significant improvement over our current choice for the joint distribution or the rest of our results. In Figure \ref{joint_distri}, we have also shown the $\avg{\fcoll | \delta}$ values for various $\delta$ bins using short horizontal black lines. The variable length of the horizontal line reflects the variable bin widths in $\delta$. We distinguish between the collapse fraction \fcoll\ computed from equation \ref{fcoll_eqn} (constrained to be between 0 and 1) and the mass-averaged collapse fraction $\fcollm(\mathbf{x}) \equiv \fcoll(\mathbf{x})(1+\delta(\mathbf{x}))$, where as usual, for the predicted (true) \fcollm the $\delta$ is taken to be from the LB (RB). We use the following expressions to compute the auto power spectrum of a field $g(\mathbf{x})$, denoted by $P_g(k)$, and its cross power spectrum with another field $h(\mathbf{x})$, denoted by $P_{gh}(k)$: 
\begin{align}\label{Pk_eqn}
\frac{\avg{g(\mathbf{k})g^*(\mathbf{k}')}}{\bar g^2} &= (2\pi)^3P_g(k)\delta_D(\mathbf{k}-\mathbf{k}') \,,\\
\frac{\avg{g(\mathbf{k}) h^*(\mathbf{k}')}}{\bar g \bar h} &= (2\pi)^3P_{gh}(k)\delta_D(\mathbf{k}-\mathbf{k}') \,,
\end{align}
where $g(\mathbf{k})$ and $\bar g$ are, respectively, the Fourier conjugate and mean of $g(\mathbf{x})$, $\delta_D$ denotes the Dirac delta function, an asterisk denotes complex conjugation and the angular brackets represent an average over Fourier space such that $|\mathbf{k}| = k$.

\begin{figure}[h]
    \centering
    \begin{subfigure}{0.49\linewidth}
    \centering
    \includegraphics[width=\linewidth, trim={0 0.7em 0 0.4em},clip]{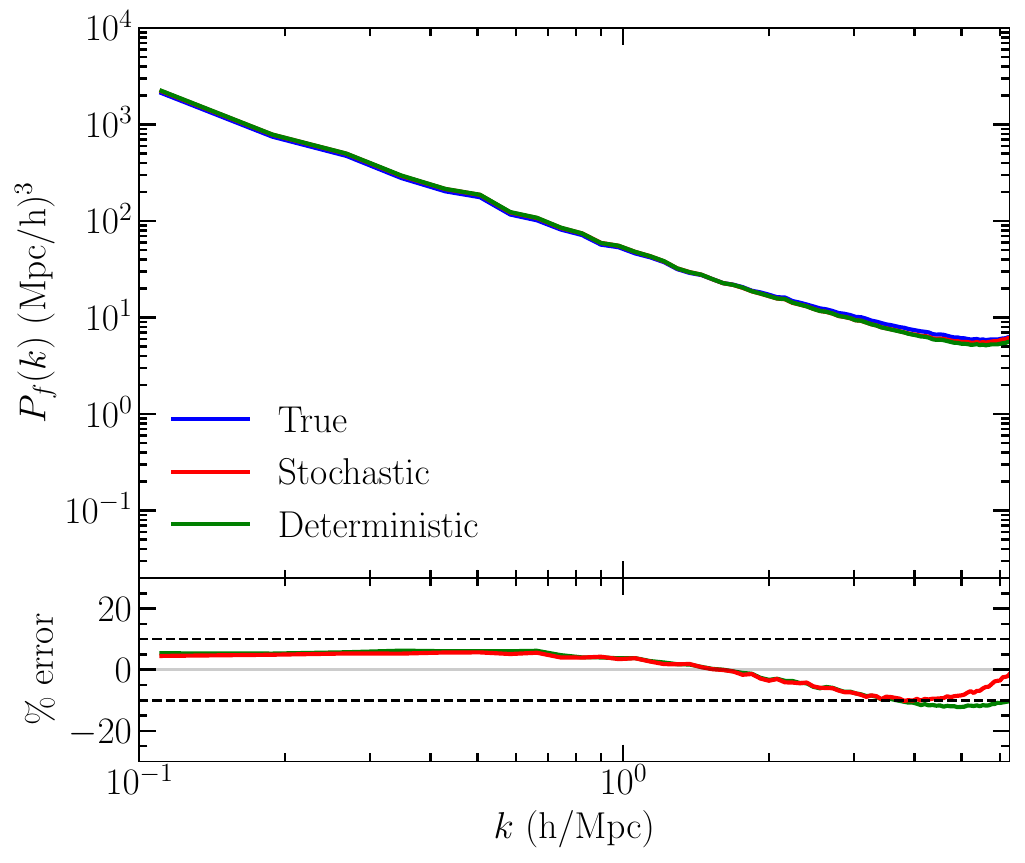}
    \caption{}
    \label{fcoll_auto_z7}
\end{subfigure}
\hspace{0em}
    \begin{subfigure}{0.49\linewidth}
    \centering
    \includegraphics[width=\linewidth, trim={0 0.7em 0 0.4em},clip]     {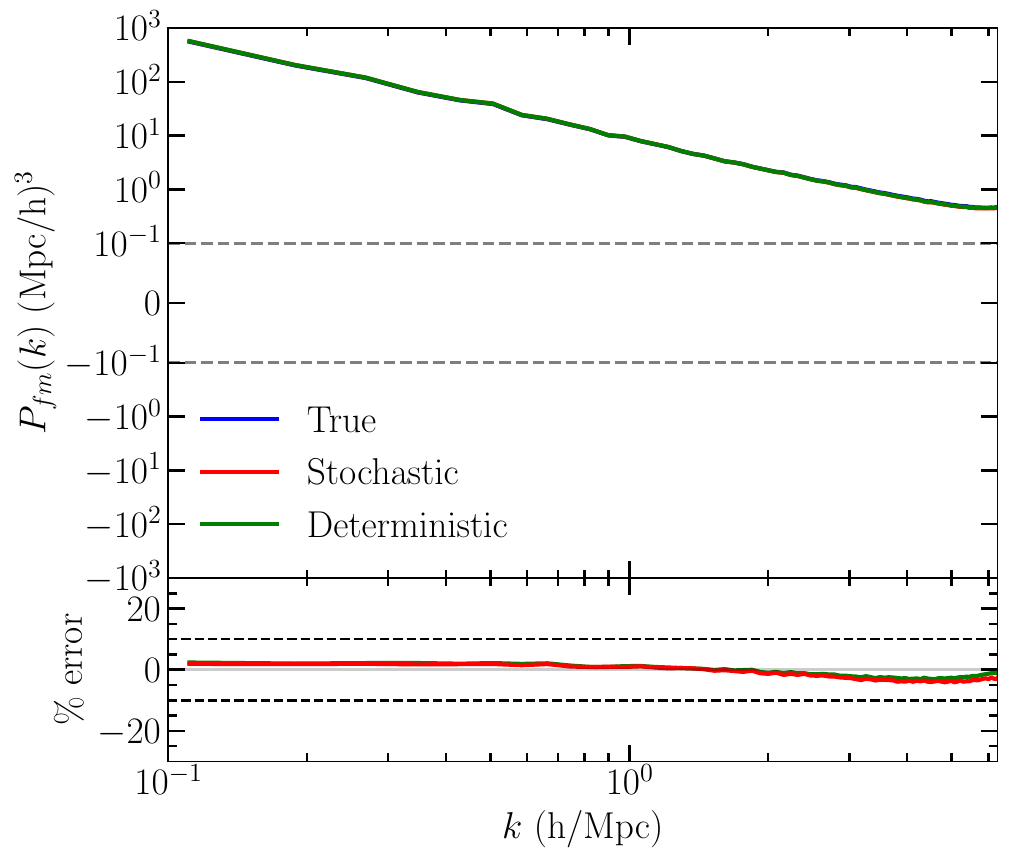}
    \caption{}
    \label{fcoll_cross_z7}
    \end{subfigure}
    \vspace{-1em}
    \caption{\small Comparison of (a) $\fcollm$-$\fcollm$ auto and (b) $\fcollm$-$\Delta$ cross power spectra, between truth and predictions using the deterministic and stochastic cases. The two cases behave similarly, having a roughly constant error of 5\% for $k < 2\ \hMpc$ in the \fcollm auto. The stochastic case has slightly more power at the smallest scales of $k \gtrsim 4\ \hMpc$, and this effect gets amplified in the corresponding HI map (see Figure \ref{z7_fid_nostoch_script}).}
    \label{fcoll_PS_z7}
\end{figure}

We compute the auto and cross power spectra by setting $g = \fcollm$ and $h = \Delta$ respectively in the above, for both the deterministic and stochastic cases, and compare them with the truth in Figure \ref{fcoll_PS_z7}. The agreement between the auto power spectra is within 5\% for $k \lesssim 2$ \hMpc, and at the smallest scales stays within 10\% for the stochastic case whereas for the deterministic case it worsens to slightly below $-10$\%. This implies that there is some extra small-scale power in the stochastic \fcoll\ field, and this difference will become more stark once we go to the HI density field in \ref{subsubsec:script results}. The cross-power spectrum is recovered better as expected, with sub-2\% errors for most of the $k$ range and only becoming $\sim 5$\% at the smallest scales. We can see that the level of agreement is very similar between the stochastic and deterministic cases.

If we take a closer look at Figure \ref{fcoll_auto_z7}, the error in the large-scale power is mostly constant for $k \leq 0.7~\hMpc$. Moreover, this error arises predominantly due to the error in the mean of \fcollm between the truth and predictions. This implies a good agreement $(\lesssim 1\%)$ at large-scales between the un-normalized power spectra, computed by dropping the $\Bar{g}^2$ in the auto power as defined in equation \ref{Pk_eqn}. We address this issue in Appendix \ref{appendix:fcoll norm error}.


\subsubsection{\textsc{script} Results}\label{subsubsec:script results}

As mentioned earlier, we use \textsc{script} to model the HI and HII fields relevant to EoR. \textsc{script} constructs ionized bubbles around sources by allowing regions to receive photons from multiple sources and possibly get `overionized' with an ionization fraction greater than 1. These excess photons are then distributed in the nearby cells causing them to become ionized, until the ionization levels of all overionized cells have been properly adjusted. In this manner, \textsc{script} incorporates photon conservation explicitly and hence achieves a large-scale HI power spectrum that is converged with respect to the spatial resolution of the \fcoll\ and density fields.

The code requires $\fcoll(\mathbf{x})$ at the desired redshift, along with the \textit{reionization efficiency parameter} $\zeta$ as the primary inputs. $\zeta$ gives the number of ionizing photons in the inter-galactic medium per hydrogen in dark matter halos. The output is an ionization (HII) fraction field $x_{\text{HII}}(\mathbf{x})$, which can then be used to get an HI fraction field, $x_{\text{HI}}(\mathbf{x}) = 1 - x_{\text{HII}}(\mathbf{x})$. Upon mass-averaging these, we get the HII and HI density fields upto normalization:
\begin{align}
x_{\text{HI}}^M(\mathbf{x}) &= x_{\text{HI}}(\mathbf{x})(1 + \delta(\mathbf{x})) \propto \rho_{\text{HI}}(\mathbf{x})\,; \\
x_{\text{HII}}^M(\mathbf{x}) &= x_{\text{HII}}(\mathbf{x})(1 + \delta(\mathbf{x})) \propto \rho_{\text{HII}}(\mathbf{x})\,.
\end{align}  

\begin{figure}[H]
    \centering
    \includegraphics[width = 0.78\linewidth, trim={0.6em 0.6em 0.6em 0.6em}, clip]{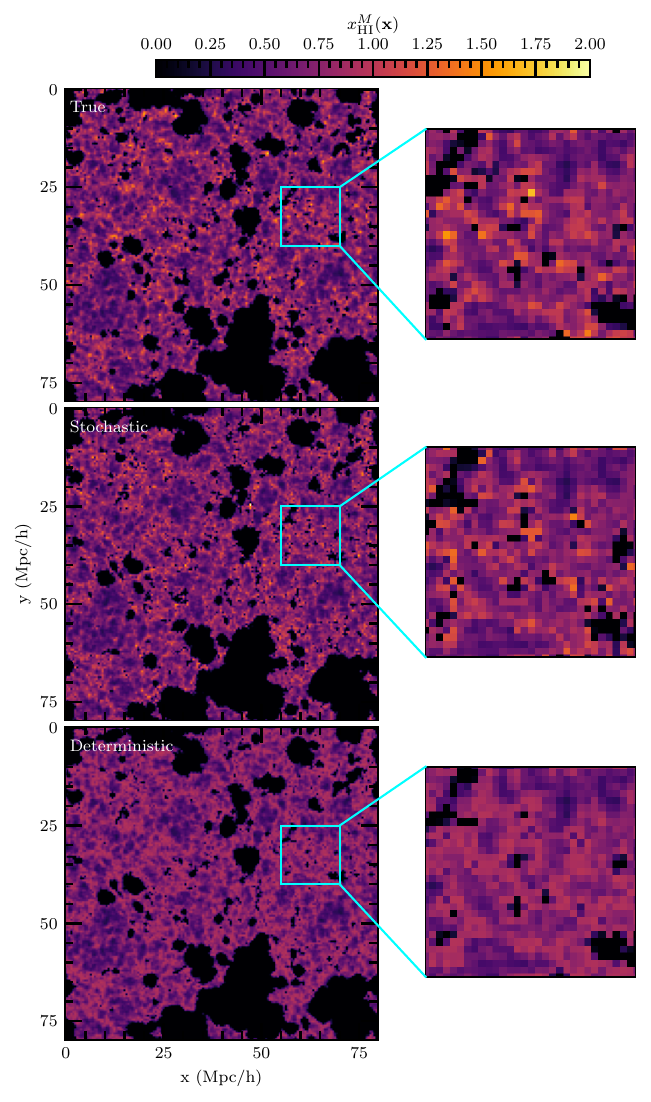}
    \caption{\small The neutral HI density field at $Q^M_{\text{HII}} = 0.5$ in the ground truth \textit{(top panel)}, as recovered by our ML interpolator (stochastic, \textit{middle panel}), and as recovered using the conditional means (deterministic, \textit{bottom panel}) at a slice through $z=50$ \Mpch. The black regions are the ionized bubbles. The deterministic case washes out small-scale HI fluctuations over a few-voxel scales to a large extent. The stochastic case recovers them better, but with some inaccuracy in the small-scale HI density correlations (refer to section \ref{sec:disc} for a detailed discussion).}
    \label{neutral_map}
\end{figure}

\begin{figure}[h]
  \centering
  \begin{subfigure}{0.49\linewidth}
    \centering
    \includegraphics[width=\linewidth, trim={0 0.7em 0 0.4em},clip]{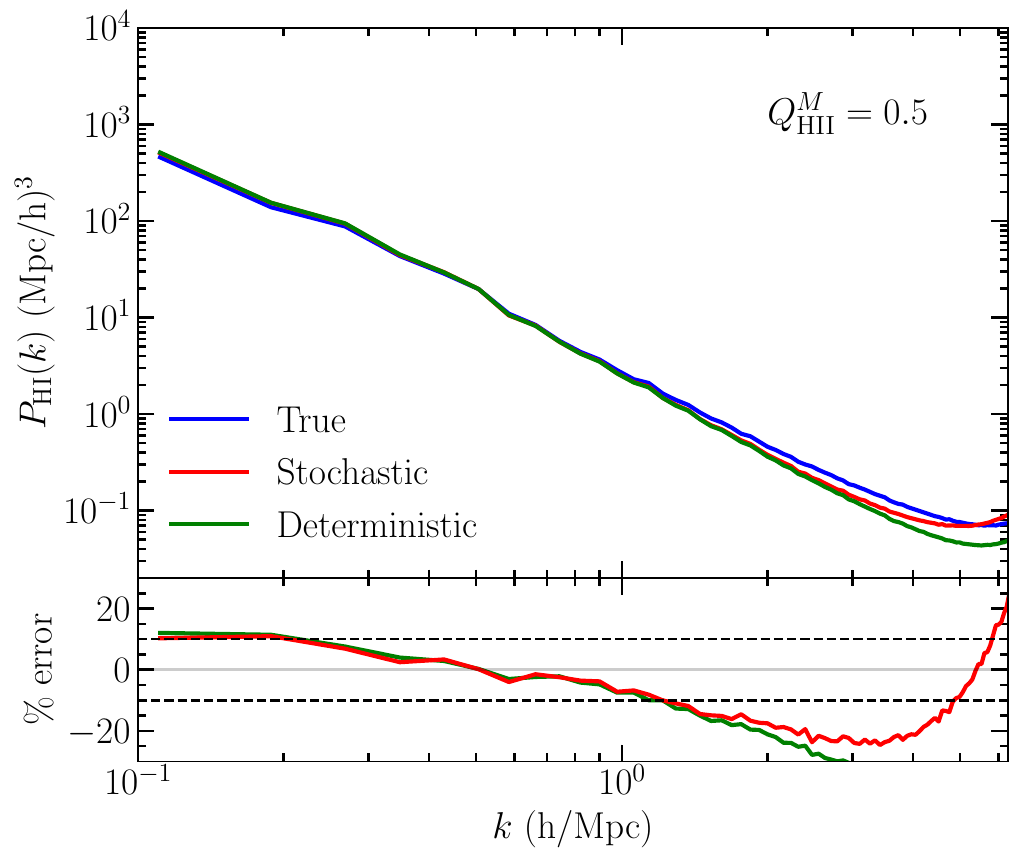}
    \caption{}
    \label{z7_HH_nostoch}
  \end{subfigure}
  \hspace{0em}
  \begin{subfigure}{0.49\linewidth}
    \includegraphics[width=\linewidth, trim={0 0.7em 0 0.4em},clip]{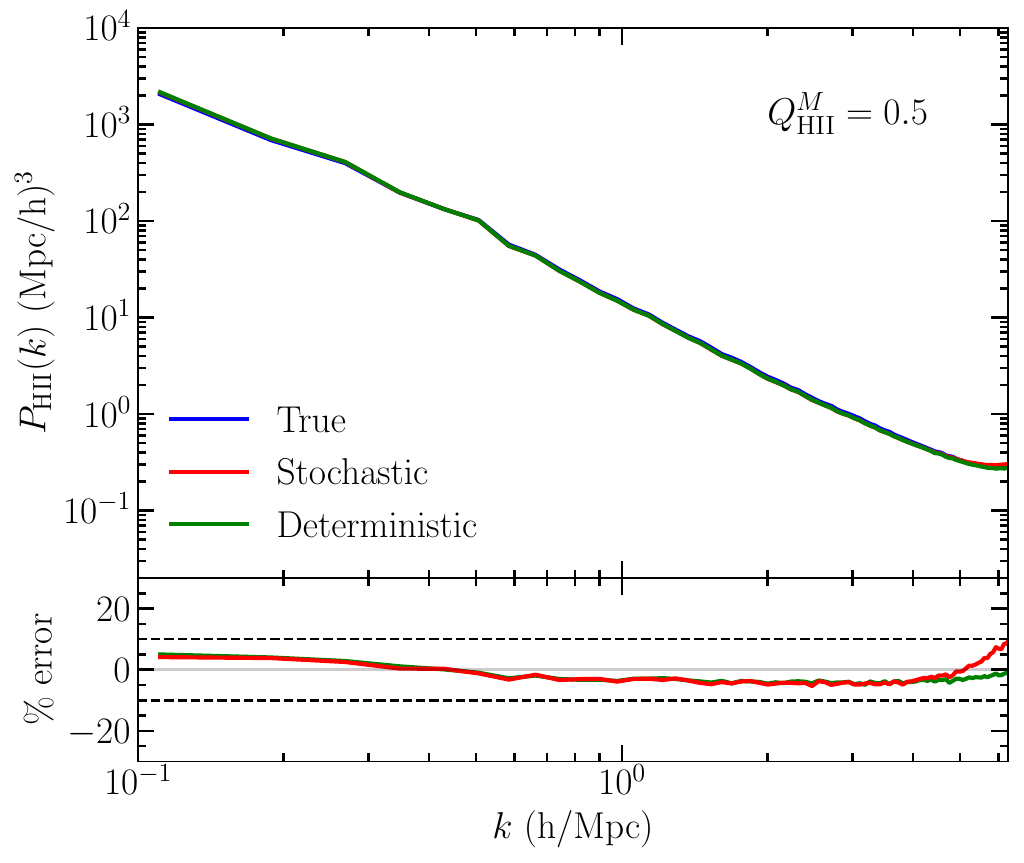}
    \caption{}
    \label{z7_QQ_nostoch}
  \end{subfigure}
  \vspace{-1em}
  \caption{\small (a) HI-HI, (b) HII-HII power spectra for truth and predictions using the stochastic and deterministic cases ($Q^M_{\text{HII}}=0.5$). In (a), the large-scale power at $k \sim 0.1\ \hMpc$ is recovered similarly at $\sim 10\%$ in both the cases, while the stochastic case has a better accuracy at small scales ($k > 2\ \hMpc$). This highlights the importance of scatter in \fcoll\ around the mean in contributing to the HI density fluctuations at small scales. In (b), the power is almost identical between the two cases across most of the $k$ range, except at very small scales ($k > 5\ \hMpc$), where the stochastic case overestimates the error to $\sim 10\%$ (see section \ref{sec:disc} for a discussion). }
  \label{z7_fid_nostoch_script}
\end{figure}

We use the stochastic, deterministic and true collapse fraction fields as the input to \textsc{script} and generate the HI and HII maps. We assume a constant ionizing efficiency $\zeta$ and calibrate it for all the three cases separately such that the global ionization fraction, $Q_{\text{HII}}^M \equiv \langle x_{\text{HII}}^M(\mathbf{x}) \rangle$ is 0.5 (this is our fiducial setting). A comparison of the HI density field $x_{\text{HI}}^M(\mathbf{x})$, at a slice through 
$z=50~ \Mpch$ is then shown in Figure \ref{neutral_map}. We also compute statistics such as the auto and cross (with $\Delta$) power spectra of the HII and HI density fields, computed as given in equation \ref{Pk_eqn}. The comparison between the deterministic, stochastic and true cases for the fiducial $Q^M_{\text{HII}} = 0.5$ can be seen in Figure \ref{z7_fid_nostoch_script}.

For the HI auto power spectrum, it is clear that the error in the recovery of large-scale power $(k \lesssim 1~\hMpc)$ is similar between the deterministic and stochastic cases, with both being atmost 10\% in magnitude. Interestingly, at the smallest scales, the deterministic case underestimates the power with a large error of around $35-40\%$ whereas the stochastic case has a better agreement of around $20-25\%$. For the HII auto power, the recovery is more consistent between the two cases, being well within 10\% for the entire $k$ range. This highlights the crucial role played by stochasticity in correctly predicting specifically the HI map during reionization, and we discuss this further in section \ref{sec:disc}.


\subsection{Variation}

\subsubsection{Semi-Analytical Methods}

The extended Press-Schechter formalism \cite{PS_74, BCEK_91} calculates the conditional mass function of dark matter halos by using an excursion set approach for identifying gravitationally collapsed regions, with a constant barrier given by the threshold linear density for spherical collapse $\delta_c$. A better match to N-body simulations is provided by the conditional mass function calculated by Sheth \& Tormen \cite{ST_99, ST_02} by accounting for ellipsoidal collapse. This leads to a barrier definition that depends on the variance of the smoothed density field at the scale under consideration. Both of these semi-analytical methods have been commonly used in semi-numerical models of reionization \cite{Mesinger_2007, zahn_11, 21cmfast} to prescribe the $\fcoll(\mathbf{x})$ field, given the underlying density field. It is also important to note that the resultant $\fcoll(\mathbf{x}|R, \delta_0)$ is obtained using a deterministic formula in the smoothing scale $R$ and the overdensity of the region $\delta_0$. This leads to an \fcoll\ field prediction analogous to our deterministic case where the scatter around the \fcoll\ values due to varying environmental features beyond $\delta$ is neglected. We aim to use the non-linear density field of the LB as an input to obtain the conditional PS and ST collapse fraction fields and compare the results with our stochastic and deterministic cases. We convert the non-linear density $\delta_\text{NL}$ from the N-body simulation into the linear one $\delta_\text{L}$ required by the semi-analytical formulae using the spherical collapse approximation described by the following implicit relations \cite{mo_white_96},
\begin{align}
&1+\delta_{\mathrm{NL}}=\dfrac{9}{2} \dfrac{(\theta-\sin \theta)^2}{(1-\cos \theta)^3}\,, \hspace{2em} \delta_{\mathrm{L}}=\dfrac{3 \times 6^{2 / 3}}{20}(\theta-\sin \theta)^{2 / 3}\,, \text { for } \delta_{\mathrm{NL}}>0\,, \\
&1+\delta_{\mathrm{NL}}=\dfrac{9}{2} \dfrac{(\sinh \theta-\theta)^2}{(\cosh \theta-1)^3}\,, \hspace{2em} \delta_{\mathrm{L}}=-\dfrac{3 \times 6^{2 / 3}}{20}(\sinh \theta-\theta)^{2 / 3}\,, \text { for } \delta_{\mathrm{NL}}<0 \,.
\end{align}

\begin{figure}[h]
    \centering
    \begin{subfigure}{0.49\linewidth}
    \centering
    \includegraphics[width=\linewidth]{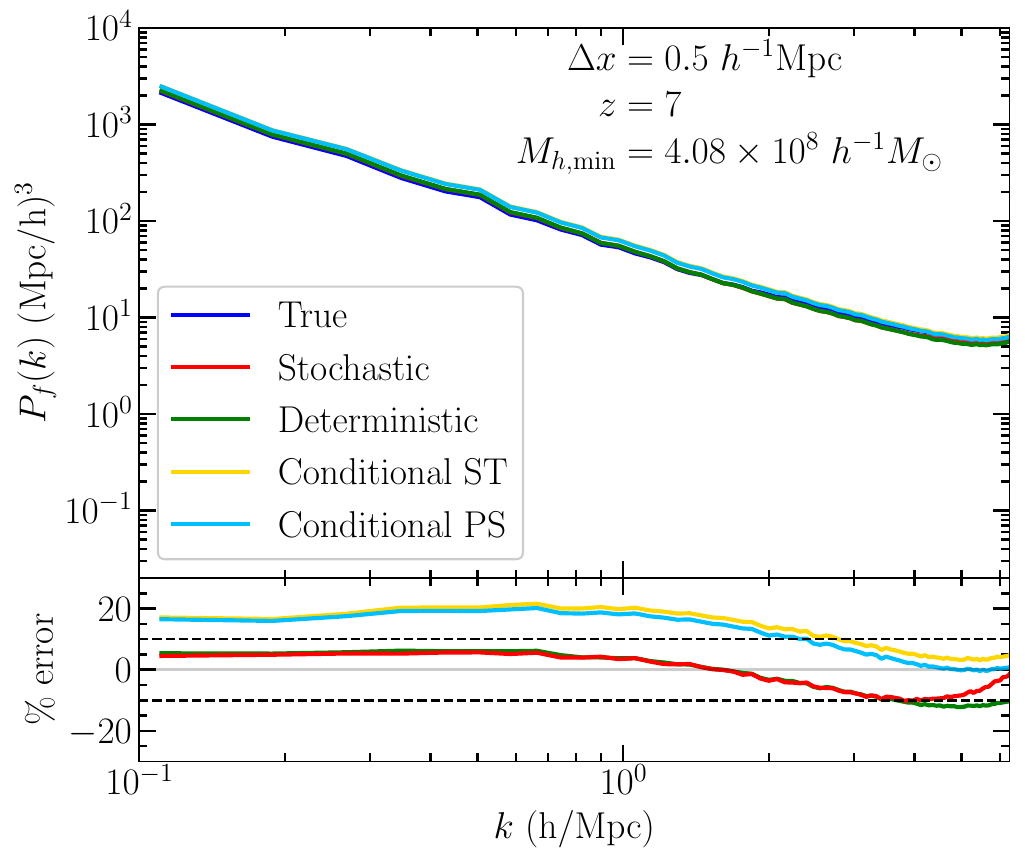}
    \caption{}
    \label{fcoll_auto_z7_full_comp}
\end{subfigure}
\hspace{0em}
    \begin{subfigure}{0.49\linewidth}
    \centering
    \includegraphics[width=\linewidth]{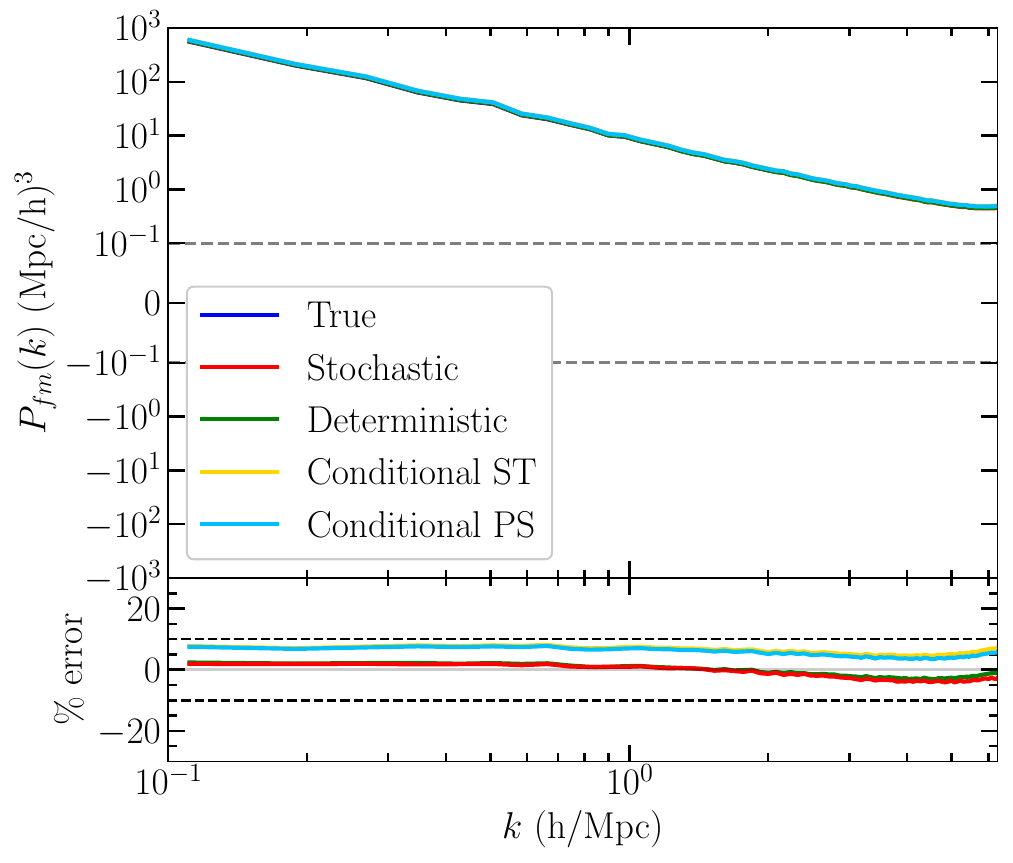}
    \caption{}
    \label{fcoll_cross_z7_full_comp}
    \end{subfigure}
    \caption{\small Comparison of (a) $\fcollm$-$\fcollm$ auto and (b) $\fcollm$-$\Delta$ cross power spectra, between truth, stochastic, deterministic and the semi-analytical predictions at the fiducial settings. The semi-analytical cases show a greater error lying between 15--20\% at large and intermediate scales ($k < 2\ \hMpc$) in the \fcollm\ power as compared to $\sim 5\%$ in the stochastic and deterministic methods over the same $k$ range. The $\fcollm$-$\Delta$ cross power is at $\sim 7$\% for the semi-analytical cases and $< 2\%$ for the deterministic and stochastic cases over similar scales. The global \fcollm mean is also relatively poorly recovered in the semi-analytical cases (see Appendix \ref{appendix:fcoll norm error}).} 
    \label{fcoll_PS_z7_full_comp}
\end{figure}

Keeping all the parameters ($Q_{\text{HII}}^M$, $z$, $M_{h, \text{min}}, \Delta x$) fixed at the fiducial\footnote{For the resolution $\Delta x=0.5\ \Mpch$, we encounter small negative \fcoll\ values that are unphysical and arise purely due to an interpolation error of the semi-analytic mass functions. They appear in cells with low overdensity and thus should be regularized to 0 before carrying out the analysis.}, we get the conditional PS and ST results by using the density field from the LB as the input. The resulting \fcollm power spectra comparison with the previously run stochastic and deterministic cases is shown in Figure \ref{fcoll_PS_z7_full_comp}. 

\begin{figure}[h]
  \centering
  \begin{subfigure}{0.49\linewidth}
    \centering
    \includegraphics[width=\linewidth]{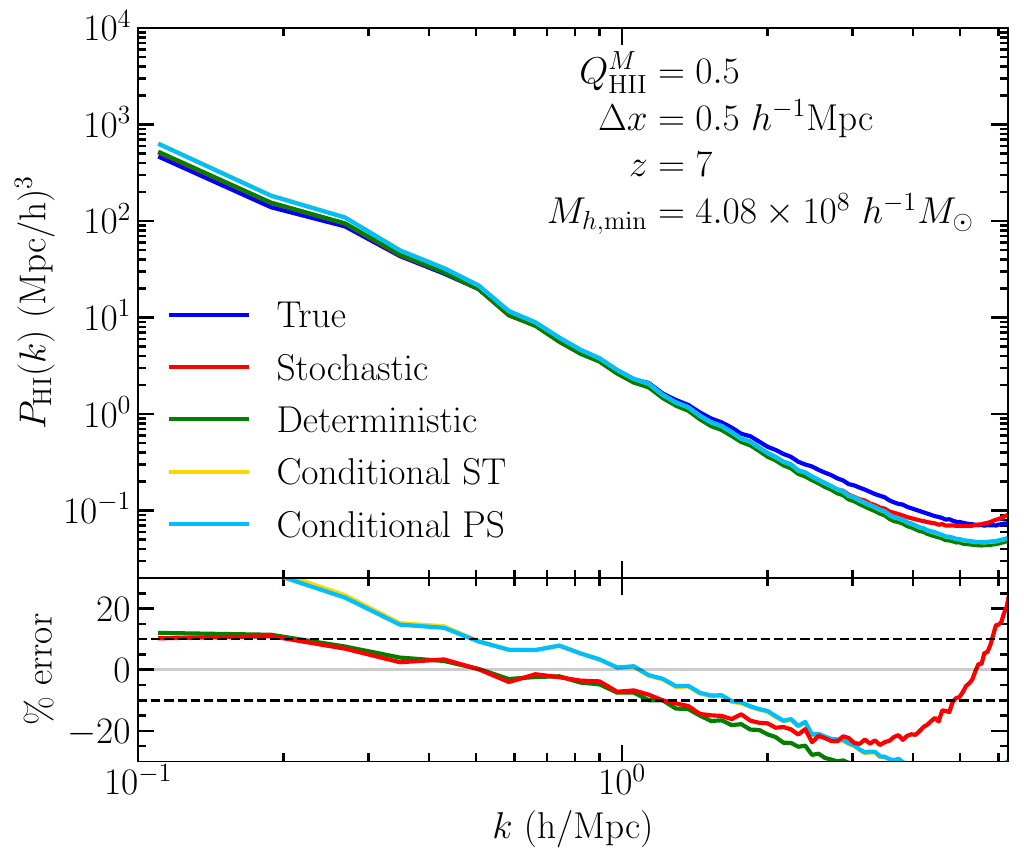}
    \label{z7_HH_full}
  \end{subfigure}
  \hspace{0em}
  \begin{subfigure}{0.49\linewidth}
    \includegraphics[width=\linewidth]{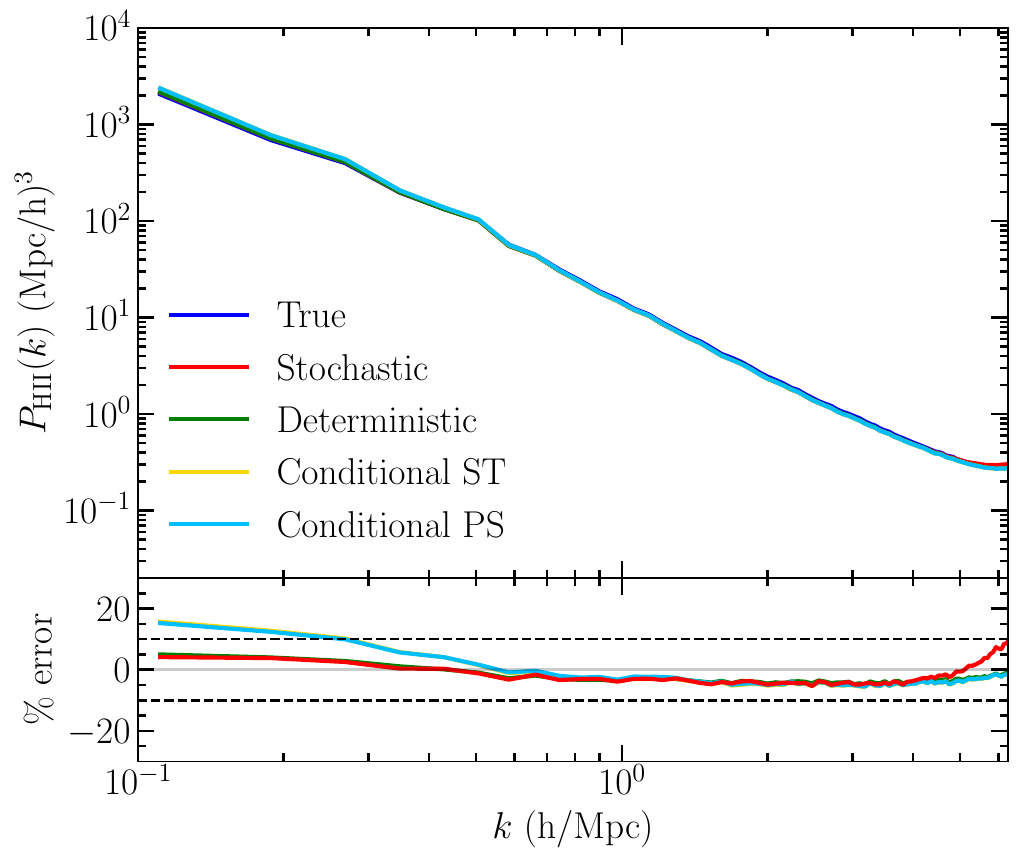}
    \label{z7_QQ_full}
  \end{subfigure}
  
  \caption{\small Comparison of HI-HI \textit{(left)}, HII-HII \textit{(right)} power spectra between truth, stochastic, deterministic and the semi-analytical predictions at $Q^M_{\text{HII}}=0.5$. For the HI power, both the semi-analytical cases show a similar error of $> 30\%$ at large scales of $k \lesssim 0.2\ \hMpc$, which is improved by the deterministic case to $\sim 12\%$ and by the stochastic case to $\sim 10\%$ at similar scales. The sign of the error flips at small scales, with the deterministic and semi-analytical cases showing $-30\%$ error or more at scales of $k \gtrsim 2\ \hMpc$, while the stochastic case being better than $-20\%$. For the HII power, at $k < 0.2\ \hMpc$, the semi-analytical cases have an error of $\sim 13\%$, which is improved to $\lesssim 4\%$ by the deterministic and stochastic cases. At small scales ($k>2\ \hMpc$), the stochastic case overestimates the error to about 10\%, unlike the rest of the cases (refer to section \ref{sec:disc}).  }
  \label{z7_script_full_comp}
\end{figure}

We can clearly observe that both the methods that use the simulations to generate \fcoll\ values (stochastic and deterministic) perform better in recovering the power than the semi-analytical prescriptions, except at very small scales. Proceeding to the HI and HII density fields and computing their power spectra, we compare the results in Figure \ref{z7_script_full_comp}. At least for the HI density field, the power at the largest and the smallest scales has a significantly greater error as compared to the stochastic case. This demonstrates the effectiveness of our method at generating accurate maps at such a fine resolution ($\Delta x = 0.5\ \Mpch$) where tidal effects become important, and the spherical collapse model used by the semi-analytical prescriptions to get the mapping between non-linear and linear density fields is inaccurate.

We now settle on the stochastic case and investigate the robustness of the method, in particular the \textsc{script} results, against a variation of the involved parameters. Hereafter, the `Predicted' label on the plots refers to the stochastic case. We perform convergence tests with respect to simulation box parameters such as grid size on the one hand, and physical parameters such as ionized fraction, redshift and minimum halo mass on the other.

\subsubsection{Physical Parameters}

\subsubsection*{Ionization Fraction}

\begin{figure}[h]
  \centering

  \begin{subfigure}{0.49\linewidth}
    \centering
    \includegraphics[width=\linewidth]{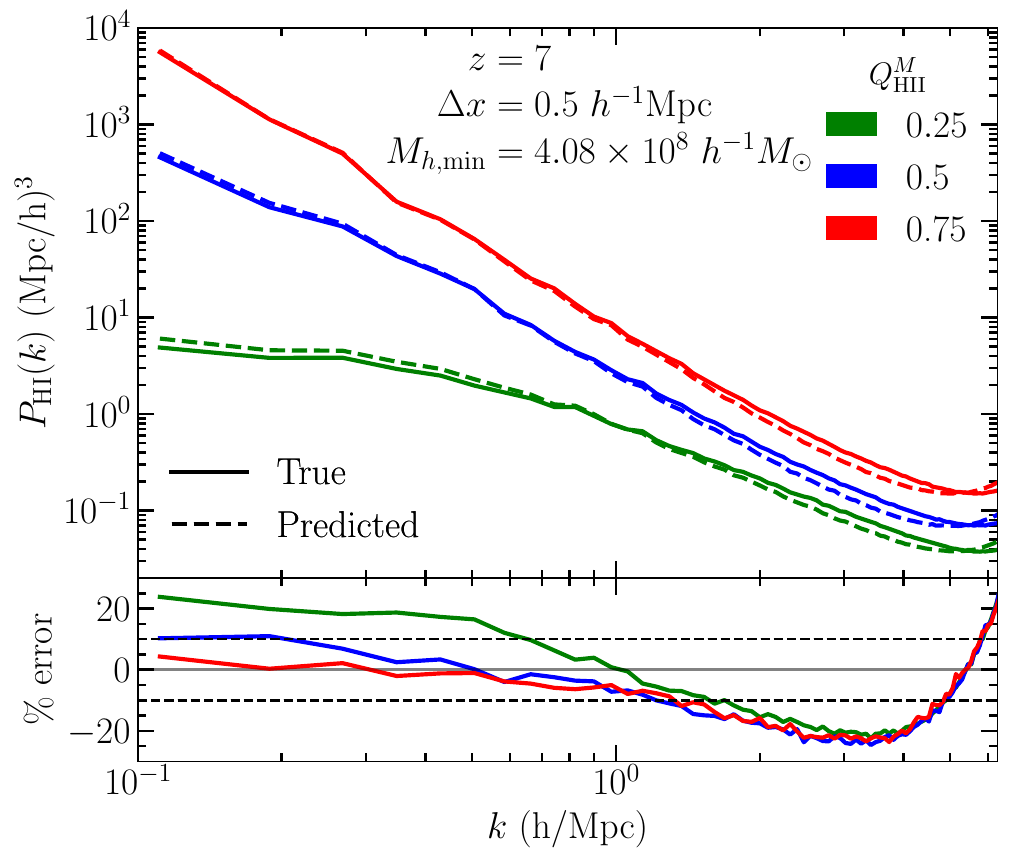}
    \label{z7_HH}
  \end{subfigure}
  \hspace{0em}
  \begin{subfigure}{0.49\linewidth}
    \centering
    \includegraphics[width=\linewidth]{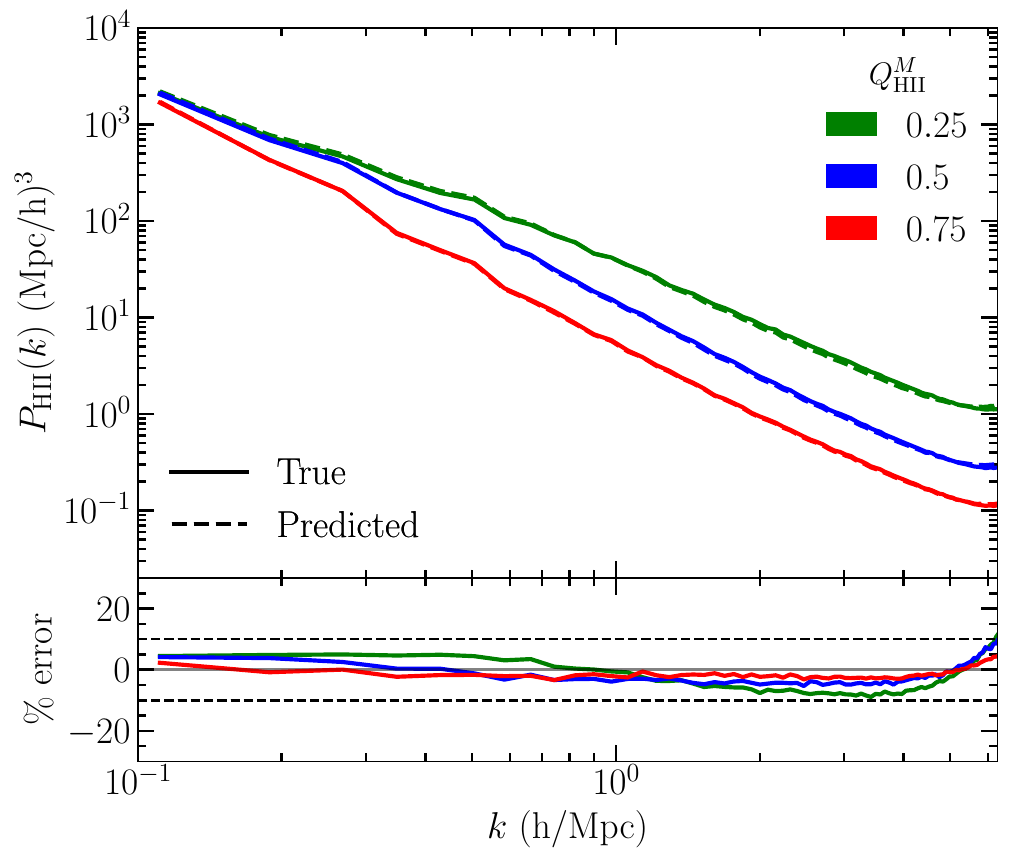}
    \label{z7_QQ}
  \end{subfigure}
  
  \vspace{-1.5em}
  
  \begin{subfigure}{0.49\linewidth}
    \centering
    \includegraphics[width=\linewidth]{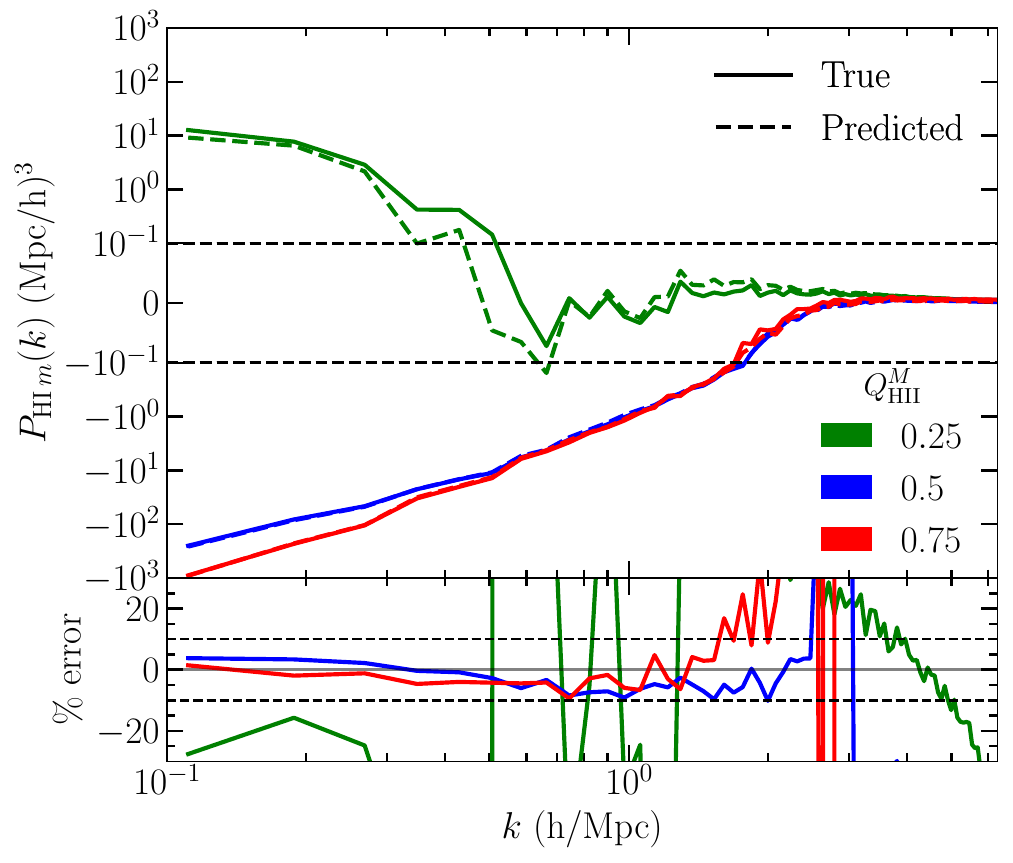}
    \label{z7_Hm}
  \end{subfigure}
  \hspace{0em}
  \begin{subfigure}{0.49\linewidth}
    \includegraphics[width=\linewidth]{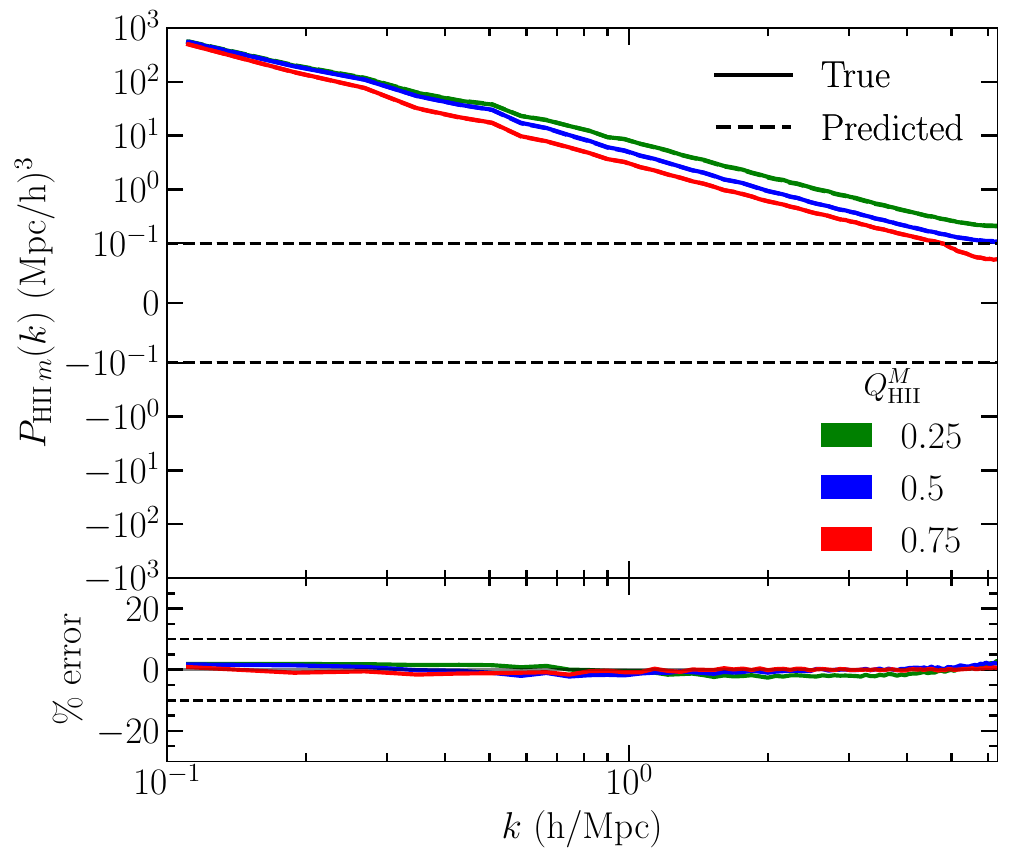}
    \label{z7_Qm}
  \end{subfigure}
  
  \caption{\small HI-HI \textit{(top left)}, HII-HII \textit{(top right)}, HI-$(1+\delta)$ \textit{(bottom left)}, HII-$(1+\delta)$ \textit{(bottom right)} power spectra for truth and prediction (always stochastic hereafter), for different values of the global ionization fraction $Q^M_{\text{HII}}=0.25, 0.5, 0.75$. We observe that the $Q^M_\text{HII}= 0.75$ case recovers the HI auto power at $\lesssim 5\%$ and the HII auto power at $\lesssim 3\%$, for large scales of $k < 1\ \hMpc$. All three cases have a similar error in the HI power at small scales of $k>1.5\ \hMpc$. The relatively large errors of $\sim 20\%$ for large scales ($k < 0.3\ \hMpc$) in the case of $Q^M_{\text{HII}}=0.25$ are discussed in section \ref{sec:disc}. }
  \label{z7_fid_script}
\end{figure}

The fraction of ionized hydrogen $Q_{\text{HII}}^M$ in the whole box is controlled by the ionizing efficiency $\zeta$. This is a crucial quantity that directly controls the size of the ionized bubbles and, hence, the evolution of the IGM during reionization. It is commonly treated as a free parameter in studies of reionization, and therefore, it is important for our method to work well for a range of $\zeta$ values. So far, we have used a $\zeta$ for the stochastic case which gives $Q_{\text{HII}}^M=0.5$, and this value comes out to be $\sim 10.8$. We now change $Q_{\text{HII}}^M$ to 0.25 and 0.75, keeping everything else the same $(M_{h, \text{ min}}=4.08 \times 10^8 ~\Mh, \Delta x=0.5 ~\Mpch, z=7)$. Figure \ref{z7_fid_script} shows the results for the auto and cross power spectra of the HI and HII density fields. 

As seen before, the large-scale HI auto power is recovered at the $\sim 10$\% level for the $Q^M_{\text{HII}}=0.5$ case, down to $k \sim 1.5$ \hMpc. The $Q^M_{\text{HII}}=0.75$ case is even better, with a 5\% error over a similar $k$ range. The HI cross power is also similar, with sub-5\% errors initially that increase to around 10\% by $k \sim 1.5$ \hMpc. The ionization field auto and cross are recovered much better, with at least a fidelity of $\sim$ 5\% down to $k \sim 2$ \hMpc, regardless of the $Q^M_{\text{HII}}$ value. We see relatively larger errors in the $Q^M_{\text{HII}}=0.25$ case and at small scales ($k \gtrsim 2$ \hMpc) even for other ionized fractions, at least in the HI results. The relatively greater disagreement for $Q^M_{\text{HII}}=0.25$ at large-scales is related to the behaviour of the large-scale HI bias (defined below) in the truth at ionization fractions close to 0.25, and is described in section \ref{sec:disc}. 

\begin{figure}[h]
  \centering
  \begin{subfigure}{0.49\linewidth}
    \centering
    \includegraphics[width=\linewidth]{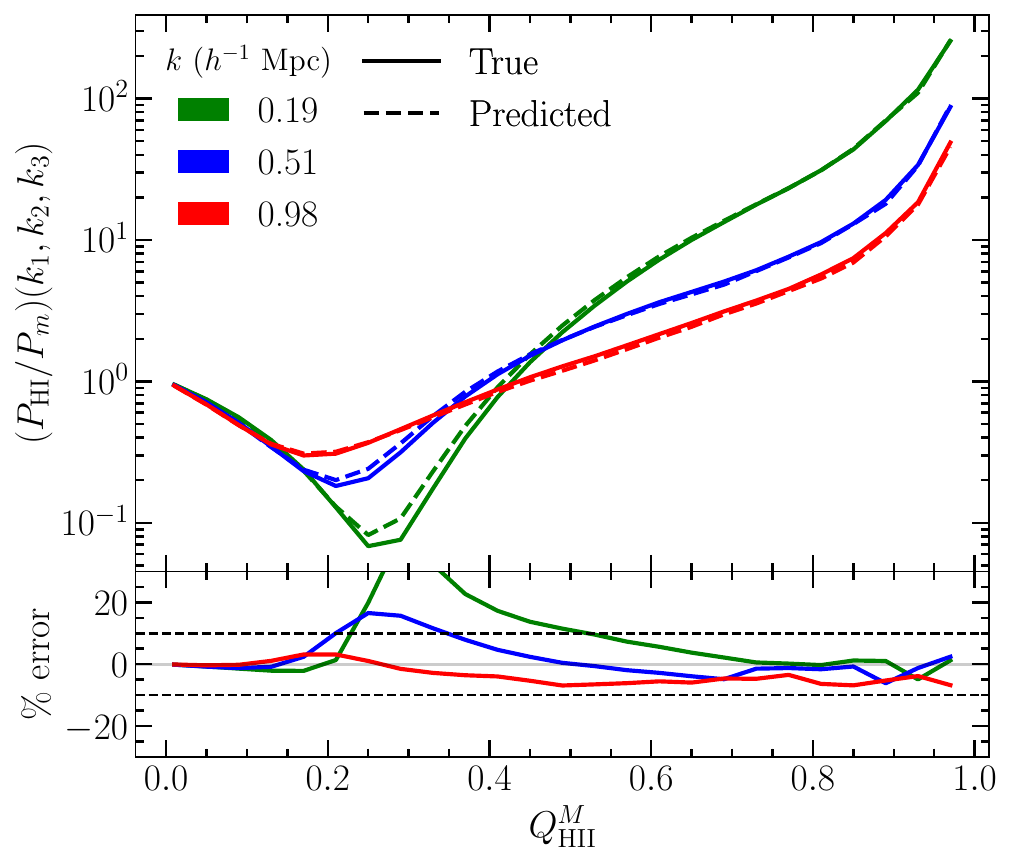}
    \caption{}
    \label{bias_HI}
  \end{subfigure}
  \hspace{0em}
  \begin{subfigure}{0.49\linewidth}
    \includegraphics[width=\linewidth]{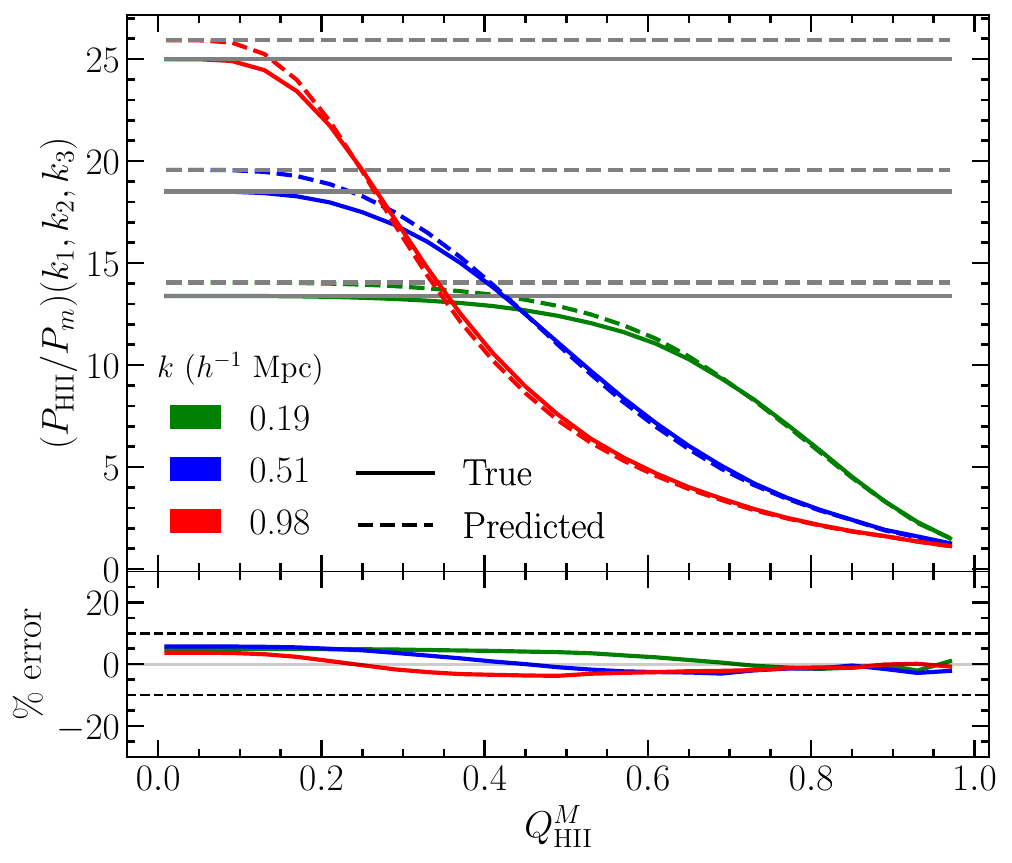}
    \caption{}
    \label{bias_Q}    
\end{subfigure}
  
  \caption{\small Comparison between truth and prediction of (a) HI bias, (b) HII bias evaluated for three different low $k$ values as a function of the ionized fraction. In the right panel, the three sets of gray horizontal lines represent the \fcoll\ bias for each of the three $k$ values. The HII bias approaches the \fcoll\ bias at large scales during sufficiently early stages of reionization, and this can be used to study the large-scale HI bias error at $Q^M_{\text{HII}}=0.25$ (see section \ref{sec:disc}). }
  \label{bias_plots}
\end{figure}

We then proceed to calculate the HI (HII) bias denoted by $b_{\text{HI}}~(b_{\text{HII}})$ and given by the expressions
\begin{equation}\label{bias_eqn}
    b^2_{\text{HI}}(k) = \dfrac{P_{\text{HI}}(k)}{P_m(k)}\,; \hspace{3em} b^2_{\text{HII}}(k) = \dfrac{P_{\text{HII}}(k)}{P_m(k)},
\end{equation}
where $P_{\text{HI}}(k)$ ($P_{\text{HII}}(k)$) is the HI (HII) auto power spectrum and $P_m(k)$ is the matter power spectrum, both computed using equation \ref{Pk_eqn}. We compute the HI and HII bias only at three different low $k$ values, and study their variation as a function of the global ionized fraction $Q^M_{\text{HII}}$ in Figure \ref{bias_plots}. In the HII bias plot, we have also plotted the \fcoll\ bias (which is independent of $Q^M_{\text{HII}}$ by construction). One can observe the HII bias to be clearly approaching the \fcoll\ bias, at sufficiently low $Q^M_{\text{HII}}$ \cite{script}. We can also see that for higher $k$, the deviation from the \fcoll\ bias happens for a lower value of $Q^M_{\text{HII}}$.


\subsubsection*{Redshift}

\begin{figure}[h]
  \centering
  \begin{subfigure}{0.49\linewidth}
    \centering
    \includegraphics[width=\linewidth]{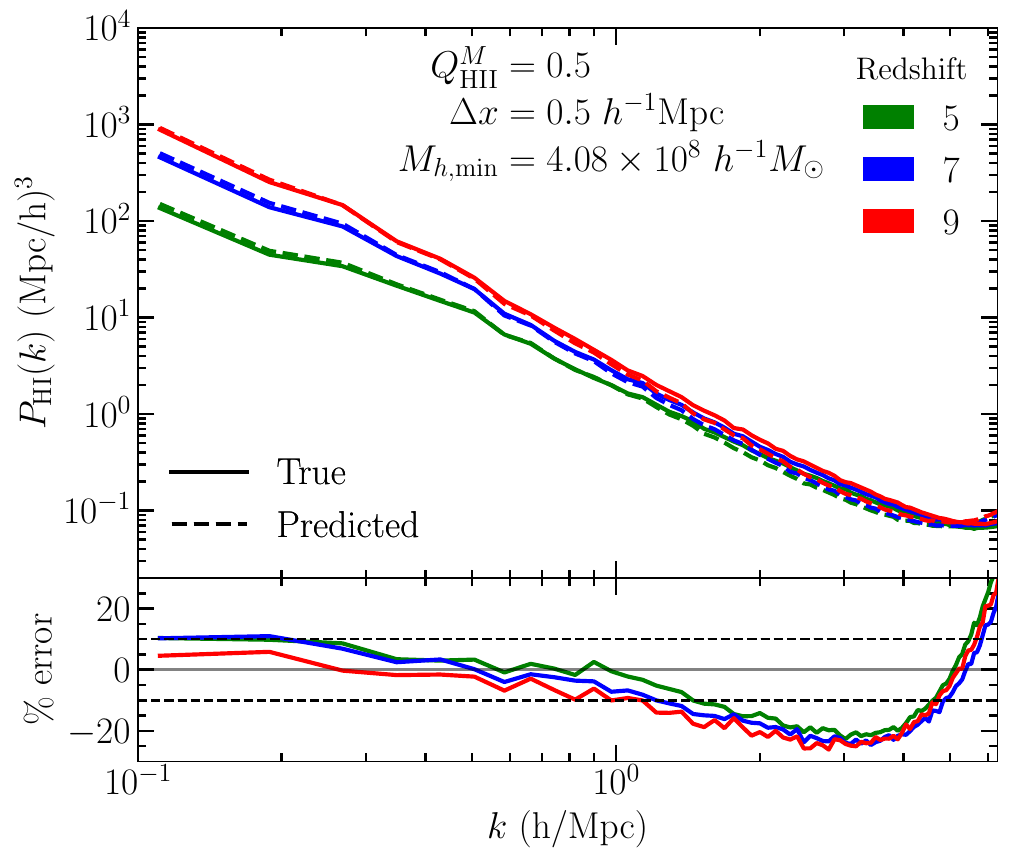}
    \label{z_var_HI}
  \end{subfigure}
  \hspace{0em}
  \begin{subfigure}{0.49\linewidth}
    \includegraphics[width=\linewidth]{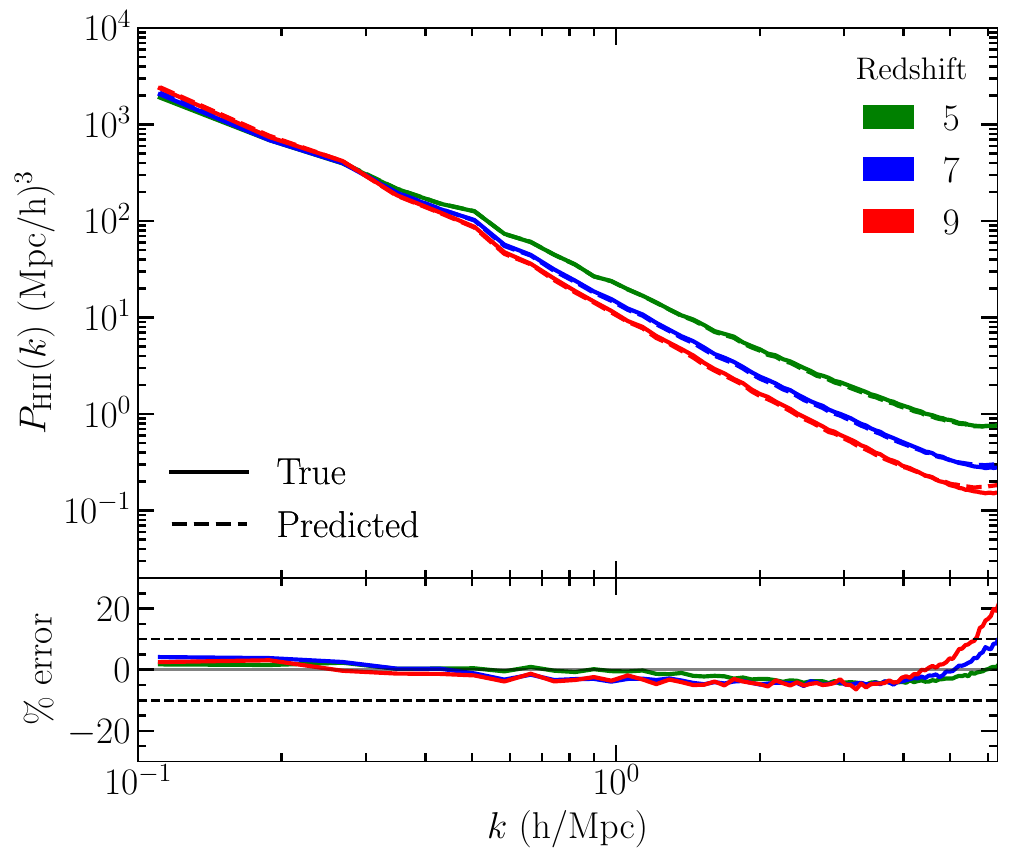}
    \label{z_var_Q}    
  \end{subfigure}

  \vspace{-1em}
  
  \caption{\small HI-HI \textit{(left panel)}, HII-HII \textit{(right panel)} power spectra for truth and prediction, for three different values of redshift at fixed ionization fraction $Q^M_{\text{HII}}=0.5$. In the HI power, $z=5$ shows a similar fidelity of 10\% and $z=9$ shows a better accuracy of $\sim 5\%$ as compared to the fiducal case, at large scales ($k \lesssim 0.2\ \hMpc$). All three redshifts share a similar error at small scales ($k \gtrsim 2\ \hMpc$). In the HII power, for most of the $k$ range, the three redshift cases have a similar error of $\lesssim 5\%$. This shows the robustness of our method for variations in redshift, which can be important in using it to study the redshift evolution of reionization. }
  \label{z_var}
\end{figure}

The redshift $z$ directly affects structure formation, with an increasing number of collapsed halos forming at lower redshifts. This changes the distribution and abundance of the sources of ionizing photons, leading to different ionized bubble topologies for a given $Q_{\text{HII}}^M$. Therefore, if we wish to use our emulator for studying the redshift evolution of the HI density field, we must ensure that it is accurate for multiple redshifts. We vary the redshift to two other values $z=5$ and $z=9$, while keeping a fixed $Q_{\text{HII}}^M = 0.5$ and $\Delta x=0.5\ \Mpch$. The effects on the HI and HII power spectra are captured in Figure \ref{z_var}. For the HI field, the agreement remains within around 10\%, at least upto $k \sim 1$ \hMpc. The overall trend is similar across redshifts, with small scales around $k \sim 3~ \hMpc$ showing a significant dip in the power. The HII results are a lot better with the errors not exceeding 5\% for almost the entire $k$ range.

\subsubsection*{Minimum Halo Mass}

The calculation of \fcoll\ at each cell assumes a halo mass cutoff $M_{h, \text{ min}}$, and this corresponds to the lowest mass halos that are capable of producing ionizing photons through the formation of luminous objects such as stars or galaxies. For halos where cooling of the infalling baryonic matter is governed by radiation through atomic transition lines, $M_{h, \text{ min}}$ is around $10^8 ~\Mh$ \cite{TRC22_review}. Changing $M_{h, \text{ min}}$ can significantly alter the ionizing photon budget and hence the distribution of ionized bubbles, since the low-mass halos are more abundant than very high-mass ones.

\begin{figure}[h]
  \centering
  \begin{subfigure}{0.49\linewidth}
    \centering
    \includegraphics[width=\linewidth]{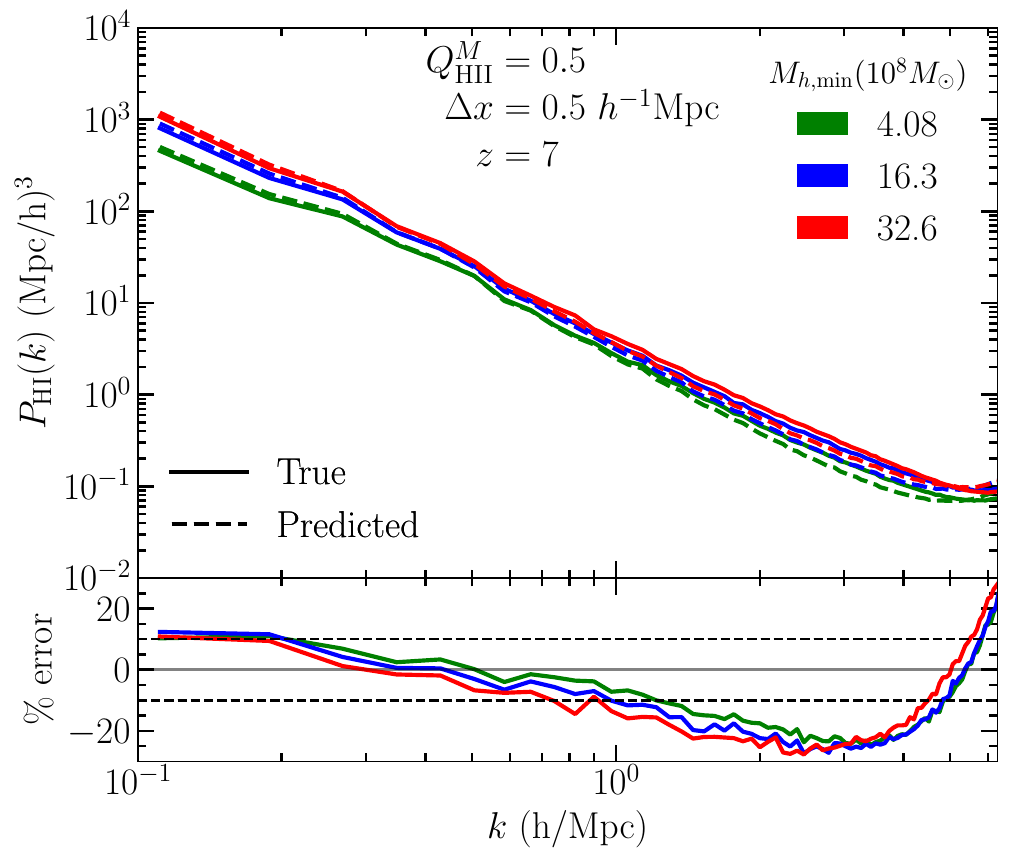}
    \label{M_min_var_HI}
  \end{subfigure}
  \hspace{0em}
  \begin{subfigure}{0.49\linewidth}
    \includegraphics[width=\linewidth]{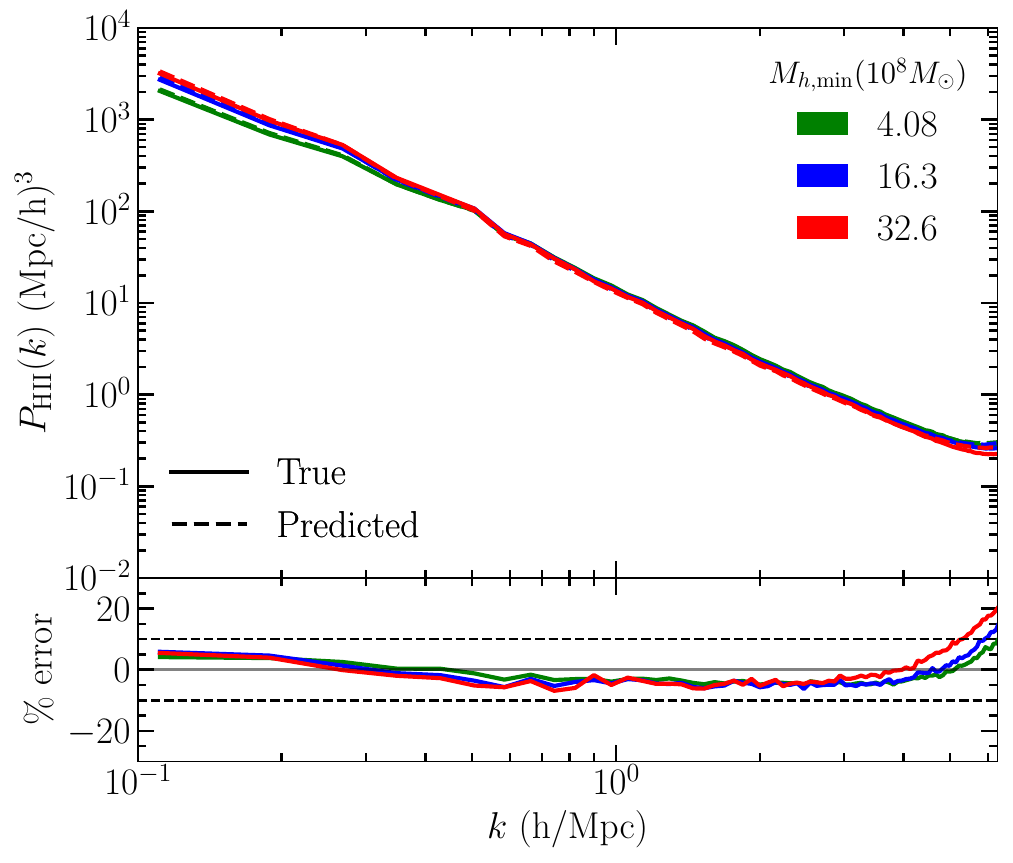}
    \label{M_min_var_Q}    
  \end{subfigure}

  \vspace{-1em}
  
  \caption{\small HI-HI \textit{(left panel)} and HII-HII \textit{(right panel)} power spectra for truth and prediction, for three different values of minimum halo mass $M_{h, \text{min}}$ at fixed ionization fraction $Q^M_{\text{HII}}=0.5$ and grid size $\Delta x=0.5$ \Mpch. The error curves are very similar for all the three cases in the HI power, with a $\sim 12\%$ error at large scales ($k < 0.2\ \hMpc$) and a $\sim -20\%$ error at small scales ($k \gtrsim 2\ \hMpc$). In the HII power, the magnitude of the error remains within around 7\% over almost the entire $k$ range. This shows that our method works with a similar fidelity for a range of $M_{h, \text{min}}$, and hence for different choices of the minimum number of particles used for identifying the FoF halos. }
  \label{M_min_var}
\end{figure}

Since different reionization models may assume different values of $M_{h, \text{ min}}$, we now vary it by varying the minimum number of particles used by the FoF group finder for identifying a halo from the default 10 to 40 and 80, while fixing $z=7$, $Q_{\text{HII}}^M = 0.5$ and $\Delta x=0.5\ \Mpch$. This changes $M_{h, \text{min}}$ from $4.08 \times 10^8$ \Mh~ to $1.63 \times 10^9$ \Mh~ and $3.26 \times 10^9$ \Mh~ respectively, and the corresponding results are shown in Figure \ref{M_min_var}. We see $\sim 12$\% error at the largest scales in the HI results, that falls and stays within 10\% till $k\sim 1$ \hMpc, and the HII power spectra remain within 7-8\% for almost the entire $k$ range.

\subsubsection{Box Parameters}

\subsection*{Grid Resolution}

\begin{figure}[h]
  \centering
  \begin{subfigure}{0.49\linewidth}
    \centering
    \includegraphics[width=\linewidth]{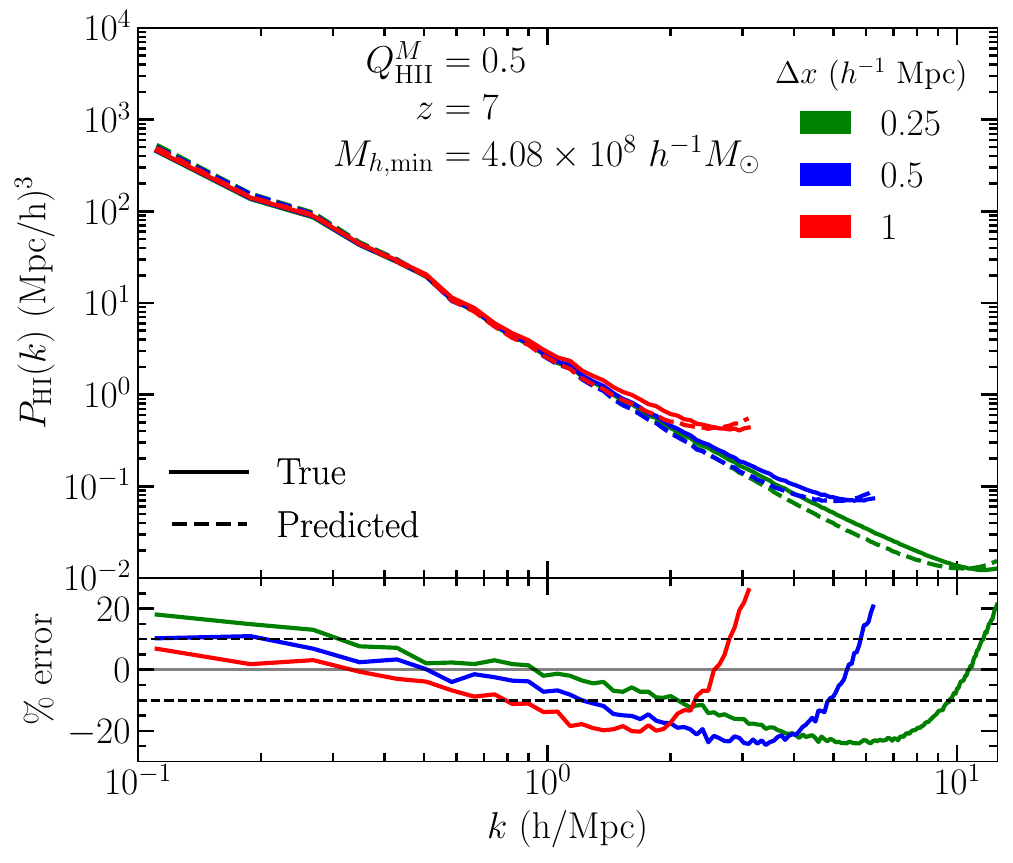}
    \label{deltax_var_HI}
  \end{subfigure}
  \hspace{0em}
  \begin{subfigure}{0.49\linewidth}
    \includegraphics[width=\linewidth]{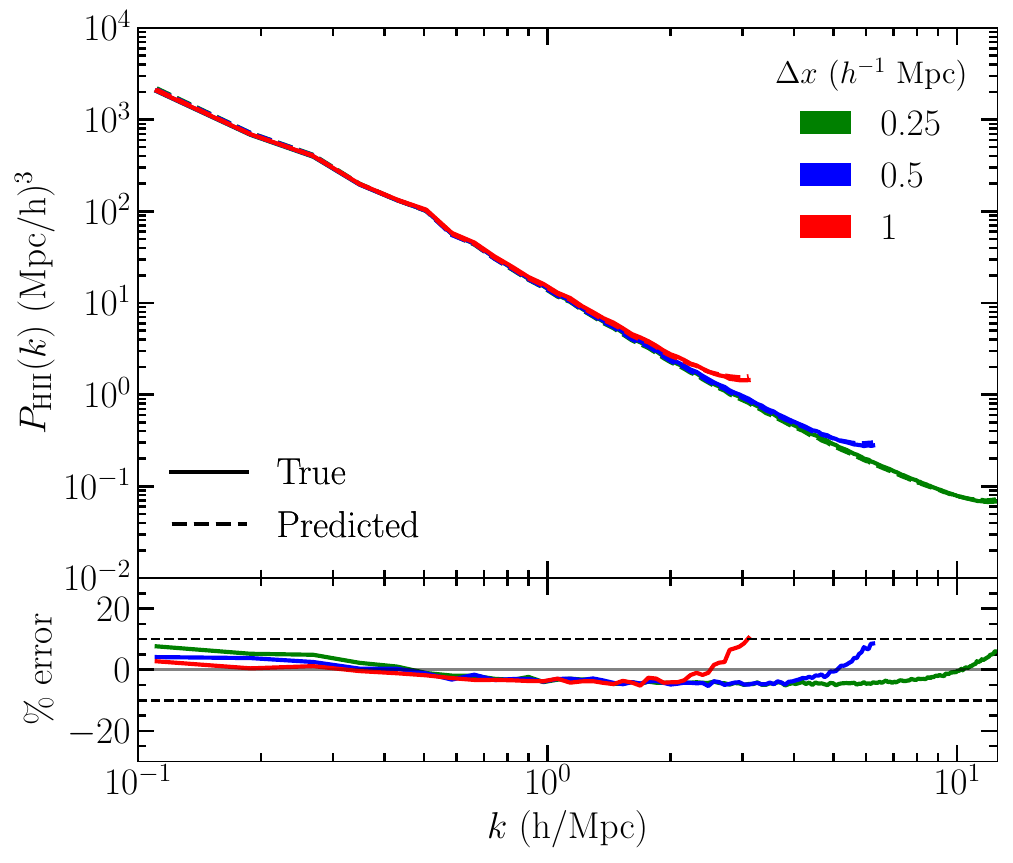}
    \label{deltax_var_Q}    
  \end{subfigure}

  \vspace{-1em}
  
  \caption{\small HI-HI \textit{(left panel)} and HII-HII \textit{(right panel)} power spectra for truth and prediction, for three different values of grid size $\Delta x$ used for getting the density and \fcoll\ fields, at fixed ionization fraction $Q^M_{\text{HII}}=0.5$. Each case has been plotted upto its Nyquist frequency. We see that the $\Delta x = 1\ \Mpch$ case has the least error in HI power ($\lesssim 5\%$) at large scales of $k < 0.5\ \hMpc$, where $\Delta x = 0.25\ \Mpch$ performs relatively poorly with $\lesssim 15\%$ error. However, the $\Delta x = 0.25\ \Mpch$ case allows us to go deeper into the $k$ range and its error crosses 10\% only at $k = 2\ \hMpc$. }
  \label{deltax_var}
\end{figure}

The grid size used for CIC-smoothing the density and \fcoll\ fields, $\Delta x$, determines the resolution at which the sources are identified and impacts the distribution of ionized regions. Rather than a physically interpretable parameter, this is a choice that one must make in order to construct the density field from the simulation box and make \fcoll\ predictions. Therefore, we check our model's sensitivity to it by varying it from the fiducial value of $\Delta x = 0.5$ \Mpch to two other values, $\Delta x = 0.25$ \Mpch and $\Delta x = 1$ \Mpch while keeping a fixed $z=7$ and $Q_{\text{HII}}^M = 0.5$, with the results shown in Figure \ref{deltax_var}. 

The $\Delta x = 0.25$ case suffers larger errors for the HI power spectra, but the other two cases have similar $\lesssim10$\% agreements at large scales below $k=1$ \hMpc. However, achieving results corresponding to $\Delta x = 0.25\ \Mpch$ is not viable using the conditional PS and ST prescriptions since one has to deal with numerical errors giving rise to negative \fcoll\ values. Our stochastic method does not suffer from such issues and represents a significant improvement over the current state of the art. The HII results are again a lot more robust, always performing better than 10\% across all $k$.

\section{Discussion}
\label{sec:disc}

Our interpolator draws \fcoll\ values for cells taking only their density information into account. This means that the correlation between sampled \fcoll\ values across different cells is controlled purely by the correlation between the density values conditioning the CDFs from which these \fcoll\ values are sampled. We expect other environmental factors to play a role in the true \fcoll\ correlation as well, but these effects are randomized across all cells by our interpolator via picking a uniform random number between 0 and 1 for inverse CDF sampling. A comparison of the recovery of \fcoll\ features by our interpolator at different scales, then, is a way of testing the sensitivity of halo formation on the cosmological environment at these scales. 

As it turns out, conditioning the \fcoll\ CDFs on the density field allows a reasonable recovery of the large-scale structure of the \fcoll\ field and consequently, the HI density map. This can be confirmed visually from the full maps in the left part of Figure \ref{neutral_map} and quantitatively through the power spectra at low $k$ in Figure \ref{fcoll_PS_z7} for the \fcoll\ field and Figure \ref{z7_fid_nostoch_script} for the HI and HII density fields. Moreover, this large-scale recovery is very similar between the stochastic and deterministic cases, with the former being marginally better. Therefore, the stochastic variations in the \fcoll\ field for a fixed matter density $\delta$, which are precisely due to the effect of other environment variables, do not affect the large-scale distribution of collapse fractions and hence the ionization bubbles within our tolerance. 

However, the \fcoll\ value in a particular cell is more strongly influenced by the neighboring density modes than density fluctuations on larger scales (say, over tens of cells). This makes the small-scale distribution of \fcoll\ values quite sensitive to the immediate environment and not just the $\delta$ value of their parent cell. Consequently, these environmental factors become more important in dictating the small-scale power of the HI density fields. If we focus on the HI density maps in Figure \ref{neutral_map}, the deterministic case completely ignores stochastic fluctuations in the \fcoll\ field and ends up producing a relatively smooth HI field outside the ionization bubbles. When we move to the stochastic case, we transition from this underlying smooth, mean-only \fcoll\ field to one that has scatter around the mean \fcoll\ incorporated into it. Apart from the correlations introduced due to the $\delta$ values, this scatter is uncorrelated cell-wise and leads to an effect similar to adding shot noise over the deterministic \fcoll. This effect gets translated to both the HI and HII maps and implies that they will have extra power in the stochastic as compared to the deterministic case. However, as explained below, there is a stronger contribution from the matter density fluctuations in the ionized bubbles to the HII power than the contribution from the weaker density field outside of these bubbles to the HI power. Both of these contributions, like the full matter power spectrum, fall with $k$. This implies that the shot noise-like contribution will be able to overcome the matter density contribution for a lower $k$ value in the HI power than in the HII power. This is evident if we note the $k$ values for which the deterministic and stochastic errors start diverging --- $k \sim 2\ \hMpc$ for HI (Figure \ref{z7_HH_nostoch}) and $k \sim 5\ \hMpc$ for HII (Figure \ref{z7_QQ_nostoch}). This can also be seen in the \fcollm\ auto power spectra itself at $k > 4\ \hMpc$ in Figure \ref{fcoll_auto_z7}.

The true case accounts for the effect of stochasticity in the correct way, increasing fluctuations in the field but doing so in a way consistent with the full information contained in the environment. One can view the \fcoll\ value at a cell as having been sampled from a Dirac delta distribution of \fcoll\ conditioned on all the cosmological environment variables that it depends on in principle, denoted by $(\delta, \alpha_1, \alpha_2, \dots)$, where $\delta$ is the dark matter overdensity as usual. All of these variables have some particular value at the cell, which dictates the \fcoll\ value. In our stochastic sampling, we are only bothering about the $\delta$ value and then uniformly sampling from the distribution conditioned only on $\delta$. This essentially amounts to assigning a `wrong' \fcoll\ that is actually associated with the variables $(\delta, \alpha_1', \alpha_2', \dots)$, which are found in some other cell. In this sense, our sampling \textit{redistributes} the \fcoll\ values from their true spatial distribution, and does so in a randomized manner, washing over structure and its correct spatial correlations at small-scales. The misplaced \fcoll\ values sampled this way that are high enough to cross the excursion-set barrier cause the corresponding cells to get flagged as ionized. This leads to the same effect in the HI density field (see the relatively more scattered and uncorrelated tiny bubbles in the middle panel of Figure \ref{neutral_map}) and thus decreases the small-scale power in the stochastic prediction as compared to the truth (Figure \ref{z7_HH_nostoch}). On the other hand, ignoring stochasticity turns out to be detrimental to the small-scale HI power of the deterministic case, leading to large errors. While our middle ground is far from the truth, it is still better at recovering the small-scale HI power than the deterministic case. 

\begin{figure}[h]
    \centering
    \includegraphics[width=\linewidth, trim={0.6em 0.6em 0.6em 0.6em}, clip]{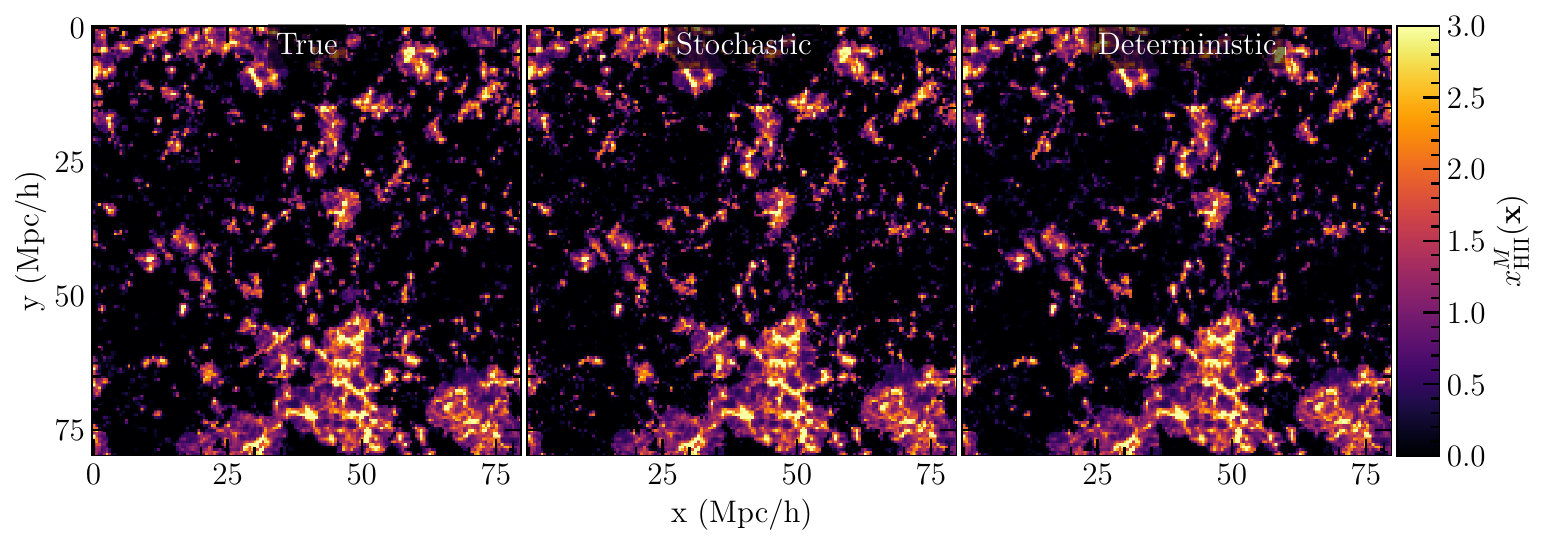}
    \caption{\small The ionized HII density field at $Q^M_{\text{HII}} = 0.5$ in the ground truth \textit{(left)}, as recovered by the stochastic \textit{(middle)} and deterministic \textit{(right)} case at a slice through $z=50$ \Mpch. The black regions contain neutral hydrogen. The low density regions are masked out and their contribution to the power spectra is subdued by the ionized bubbles that trace the high-density regions of dark matter.}
    \label{ionization_map}
\end{figure}

It is then also interesting to note the behaviour of the ionized field. Not only do the stochastic and deterministic cases recover the HII power spectra almost equally well (right panel of Figure \ref{z7_fid_nostoch_script}), the errors are significantly lesser as compared to the HI power spectra (compare left and right panels of Figure \ref{z7_fid_script}). We can understand this relative difference in the errors qualitatively as follows. As evident from Figure \ref{ionization_map}, the structure of the HII map consists of large, continuous ionized bubbles that trace out the strong regions of the density fields inside them (where the ionized fraction is 1 everywhere, i.e. there are no `holes') and numerous small ionized `islands' spread around these regions. From Figure \ref{neutral_map}, it is reasonable to expect that the differences between our predictions (stochastic or deterministic) and the truth mostly lie in how these small ionized `islands' are treated, with the large-scale ionized bubbles being recovered well. Consequently, their contribution to the HII power is not substantial against that from the strong density fluctuations inside the ionized bubbles. This leads to errors in the placement of these small ionized islands having little impact on the HII power error. The HI map, on the other hand, can be considered to be a combination of the weak regions of the matter density field (corresponding to all the regions outside the ionized bubbles) and some extra holes in it (corresponding to the small ionized islands). In such a case, the placement of these holes becomes a much more important factor in affecting the HI power since the contribution from the matter density fluctuations is not very strong.

The spurious ionized islands arise from random fluctuations in \fcoll\ that contribute in tandem to decreasing the power at small scales, but average out when large scales are considered, thereby not contributing much to the large-scale power. This is apparent from Figure \ref{fcoll_auto_z7}, where the large-scale power of the \fcollm~field has a constant offset at around 5\%. If we plot the un-normalized predicted \fcollm~auto power spectrum (that is, without dividing by $\Bar{g}^2$ in equation \ref{Pk_eqn}) then it matches the truth to within 1\%. Therefore, the observed $\sim 5\%$ at large scales is mostly due to the error in the global \fcollm~mean (squared) made by the interpolator. This arises due to a combination of binning effects, the GPR emulator error and inherent difference between the SB and RB, and is discussed further in Appendix \ref{appendix:fcoll norm error}. 

The relatively large deviation arising in the HI power at the large scales in Figure \ref{z7_fid_script} can be understood in the following manner. For the HI (HII) field, we are actually plotting the power spectrum of $\Delta_{\text{HI}} (\mathbf{x})$ ($\Delta_{\text{HII}} (\mathbf{x})$) given by
\begin{equation}
    \Delta_{\text{HI}} (\mathbf{x}) = \dfrac{x_{\text{HI}}^M(\mathbf{x})}{1-Q_{\text{HII}}^M}\,; \quad \quad \quad 
    \Delta_{\text{HII}} (\mathbf{x}) = \dfrac{x_{\text{HII}}^M(\mathbf{x})}{Q_{\text{HII}}^M}\,,
\end{equation}
and these can be related in the following manner:
\begin{align}
    \Delta_{\text{HI}} (\mathbf{x}) &= \dfrac{x_{\text{HI}}^M(\mathbf{x})}{1-Q_{\text{HII}}^M} \\
            &= \dfrac{x_{\text{HI}}(\mathbf{x})(1+\delta(\mathbf{x}))}{1-Q_{\text{HII}}^M} \\
            &= \dfrac{1+\delta(\mathbf{x}) - x_{\text{HII}}^M(\mathbf{x})}{1-Q_{\text{HII}}^M} \\
            &= \dfrac{1+\delta(\mathbf{x}) - Q_{\text{HII}}^M\Delta_{\text{HII}}(\mathbf{x})}{1-Q_{\text{HII}}^M}\,.
\end{align}
Following the discussion in Appendix B of \cite{script}, if we assume the bias to be scale-free at large scales during the early stages of reionization, we can relate the HI and HII bias as
\begin{equation}
    b_{\text{HI}} = \dfrac{1-Q_{\text{HII}}^M b_{\text{HII}}}{1-Q_{\text{HII}}^M}\,.
\end{equation}
Recall that the square of the bias is simply the power spectrum of the relevant field normalized by the matter power spectrum (equation \ref{bias_eqn}), and since the matter power spectrum at large scales is identical between LB and RB, the power spectra error is directly proportional to the bias error. We can write this at fixed $Q_{\text{HII}}^M$ as
\begin{equation}
    \dfrac{(b_{\text{HI}})_{\text{predicted}}}{(b_{\text{HI}})_{\text{true}}} = \dfrac{1-Q_{\text{HII}}^M (b_{\text{HII}})_{\text{predicted}}}{1-Q_{\text{HII}}^M (b_{\text{HII}})_{\text{true}}}\,.
\end{equation}
From Figure \ref{bias_Q}, we can read off the value of $(b^2_{\text{HII}})_{\text{true}}$ for the smallest $k$ to be around 13.5. This implies $(b_{\text{HII}})_{\text{true}} \approx 3.7$ and so the denominator in the relative error expression above will blow up around $Q_{\text{HII}}^M \approx 1/3.7 \approx 0.27$. Thus, the value of $Q_{\text{HII}}^M=0.25$ for which we plot the power spectra in Figure \ref{z7_fid_script} is also expected to show a large error. The same calculation is confirmed from Figure \ref{bias_HI} as well, where both the true and predicted HI biases become numerically very small, causing the errors to blow up. 

\section{Conclusion}
\label{sec:conc}

The advent of more advanced radio interferometer experiments such as the SKA will provide more precise bounds on the 21 cm power spectra, and hence the HI density distribution from the Epoch of Reionization (EoR). This makes the forward modeling of HI maps during EoR crucial for testing our understanding of the epoch. Efficient methods to do this require the distribution of the fraction of mass in dark matter halos (collapse fraction field) to be input into excursion-set based semi-numerical models of reionization \cite{Mesinger_2007, zahn07, choudhury09_ES, 21cmfast, lin16_ES, script}. Obtaining the collapse fraction field using the semi-analytical formalism of the conditional Press-Schechter \cite{PS_74, BCEK_91} and conditional Sheth-Tormen \cite{ST_99, ST_02} mass functions, while efficient, is an approximation to more accurate results obtained from high-dynamic range N-body simulations \cite{reed07, tinker08, courtin10, crocce11, bhattacharya11}. The latter are extremely inefficient for parameter estimation due to their high computational cost.

While there have been attempts to make the prediction of the collapse fraction field more efficient by using hybrid approaches that combine information from low-dynamic range boxes \cite{ahn12, iliev14_size, mcquinn07, santos_10, doussot22}, they have not taken into account the full stochasticity in \fcoll\ for a fixed dark matter density contrast $\delta$, as predicted by N-body simulations. In this work, we build a machine learning model to accurately predict $\fcoll(\mathbf{x})$ using a hybrid approach while taking into account the full stochasticity. We use the conditional cumulative distribution functions $\text{CDF}(\fcoll | \delta)$ obtained from a set of 7 small-volume, high-resolution simulations (SB) to train the ML model using a methodology based on Gaussian Process Regression (GPR). The density input from a large-volume, low-resolution simulation (LB) is then used to randomly draw samples of \fcoll\ values from the emulated CDFs for each cell. This constitutes our \textit{stochastic} case, and we also obtain $\fcoll(\mathbf{x})$ corresponding to the \textit{deterministic} case, which excludes stochasticity by simply using the conditional means $\avg{\fcoll | \delta}$ computed from the SB. 

Upon comparing the auto power spectra of the mass-averaged $\fcoll(\mathbf{x})$ and its cross with $\Delta \equiv 1 + \delta$ for our fiducial choice of the parameters $Q_\text{HII}^M, z, \Delta x, M_{h, \text{min}}$, we find similar levels of agreement between the stochastic and deterministic cases (Figure \ref{fcoll_PS_z7}). We then compute the HI and HII density fields using the semi-numerical code for reionization \textsc{script}. While the recovery is similar at large scales, the deterministic case performs much worse at smaller scales for the HI density field (Figure \ref{z7_fid_nostoch_script}). We then make a comparison between the simulation-based deterministic and stochastic methods and the semi-analytical conditional mass functions. For the mass-weighted \fcoll, HI and HII power spectra, the simulation-based methods work better and the stochastic case is the best at recovering the small-scale HI power (Figures \ref{fcoll_PS_z7_full_comp} and \ref{z7_script_full_comp}). We further test the flexibility of the stochastic case against variations in all the involved parameters, including global ionized fraction, redshift, grid size and minimum halo mass. For almost all the cases, we are able to recover the HI large-scale power $(k \lesssim 1\ \hMpc)$ at the $\lesssim10$\% level, whereas for the HII density field the errors are well within 10\% for the entire range of $k$ values. The accuracy, combined with its significantly lesser RAM requirements of $\sim 20$ GB for running the SB and LB as compared to $\sim 160$ GB for running the RB, makes our method a powerful tool for RAM-limited users conducting studies of reionization parameter space exploration who wish to run a single cosmological high-dynamic range simulation.

Using only the dark matter density contrast to condition the distribution of \fcoll, we are able to recover large-scale structures well in the \fcoll\ field and the subsequent HI maps. We demonstrate how stochasticity in the \fcoll\ predictions can play a critical role in recovering the small-scale structure of the HI maps. However, our specific implementation of stochasticity does not take into account the full information contained in the cosmological environment, and this leads to some spurious small-scale structures in the HI maps. Therefore, further improvements to the ML framework can include finding a set of variables, that can better reflect the environment than $\delta$ alone, to condition the distribution of \fcoll. As suggested by \cite{barsode24}, the three eigenvalues of the tidal tensor evaluated at each location $\{\lambda_1(\mathbf{x}), \lambda_2(\mathbf{x}), \lambda_3(\mathbf{x}) \}$ could be used for such a purpose, and this shall be explored in future work. The GPR machinery set up in this work will become more beneficial in this regard than a simple linear interpolation scheme, due to the high dynamic range of the 3 eigenvalues.

Another possible direction for the future entails increasing the dynamic range gap between SB/LB and RB. Currently, we are using SB simulations that are 8 times smaller in volume than the target RB. We can test the accuracy of the framework for a simulation that is 64 times smaller. One can also explore using our ML model to build a redshift evolution of reionization by sampling $\fcoll(\mathbf{x})$ at appropriately spaced redshifts. In conclusion, the method presented in this work can prove to be an efficient yet accurate way to study models of reionization and also help constrain parameters from upcoming observations. 

\acknowledgments

The authors thank the anonymous referee for an insightful review. GP thanks Susmita Adhikari and Saee Dhawalikar for valuable discussions. The research of AP is supported by the Associateship Scheme of ICTP, Trieste.  We received support from the computing facilities at NCRA for running the GADGET-2 simulations. The resources provided by the PARAM Brahma facility at IISER Pune which is a part of the National Supercomputing Mission (NSM) of the Government of India are also gratefully acknowledged. 

\section*{Data availability}

The parameters and code of the GPR emulator for the various cases can be made available upon reasonable request to the corresponding author. 

\bibliography{references}

\appendix

\section{Convergence of Results}

\begin{figure}[h]
    \centering
    \begin{subfigure}{0.49\linewidth}
    \centering
    \includegraphics[width=\linewidth]{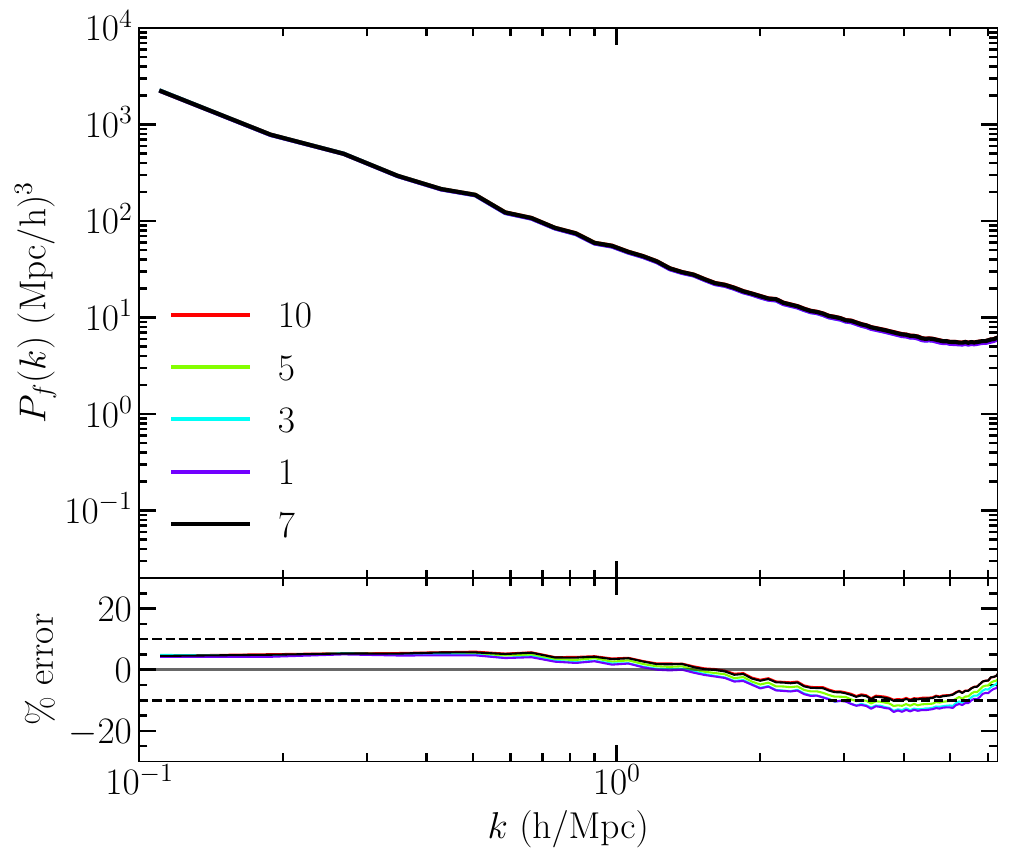}
    \caption{}
    \label{conv_fcoll_auto}
\end{subfigure}
\hspace{0em}
    \begin{subfigure}{0.49\linewidth}
    \centering
    \includegraphics[width=\linewidth]     {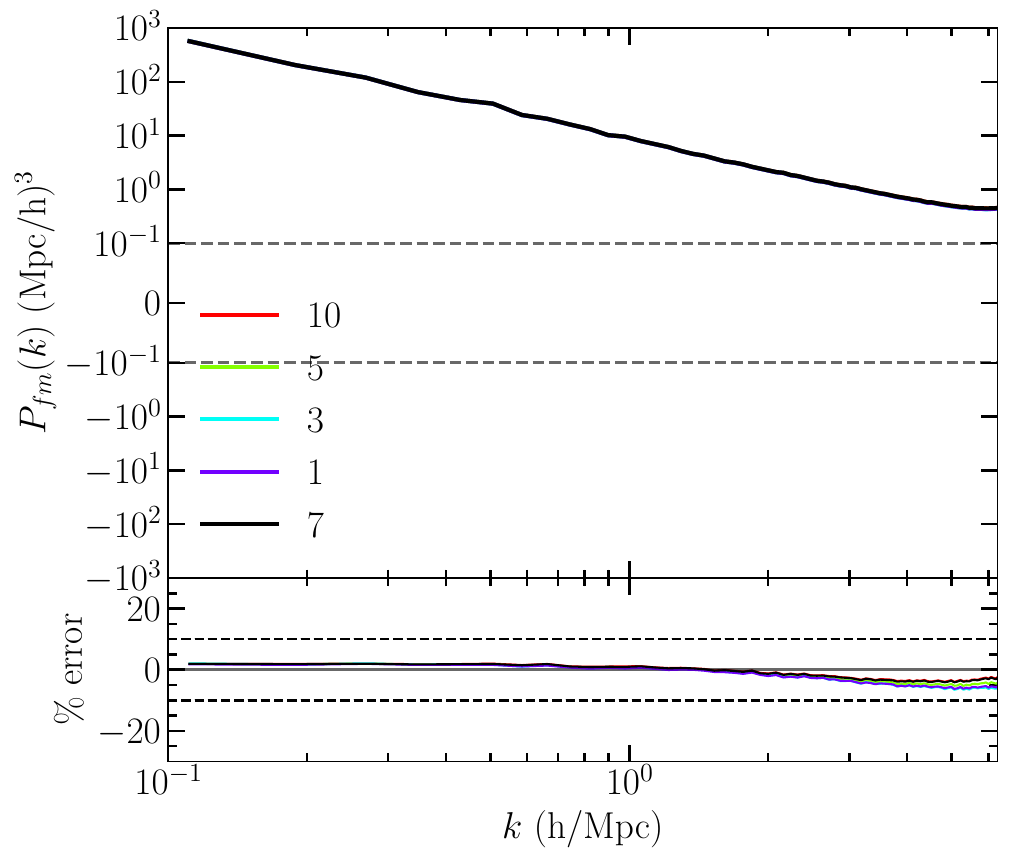}
    \caption{}
    \label{conv_fcoll_cross}
    \end{subfigure}
    
    \caption{\small Comparison of (a) $\fcollm$-$\fcollm$ auto and (b) $\fcollm$-$\Delta$ cross power spectra (upper panels) for different numbers of SB boxes combined for training, and the relative error of each with the true power spectra (lower panel). The default case that we work with is 7, shown in black. The differences between the errors are very small and mostly visible at small scales.}
    \label{conv_fcoll}
\end{figure}

\begin{figure}[h]
  \centering
  \begin{subfigure}{0.49\linewidth}
    \centering
    \includegraphics[width=\linewidth]{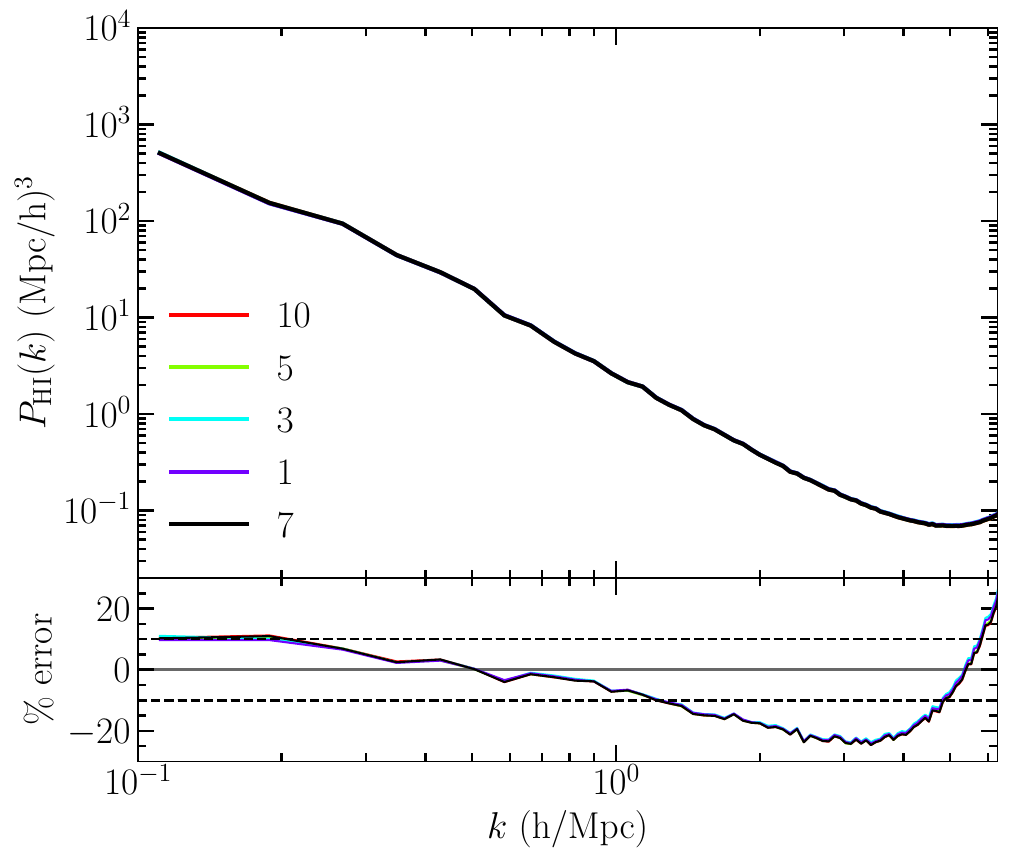}
    \label{conv_script_HI}
  \end{subfigure}
  \hspace{0em}
  \begin{subfigure}{0.49\linewidth}
    \includegraphics[width=\linewidth]{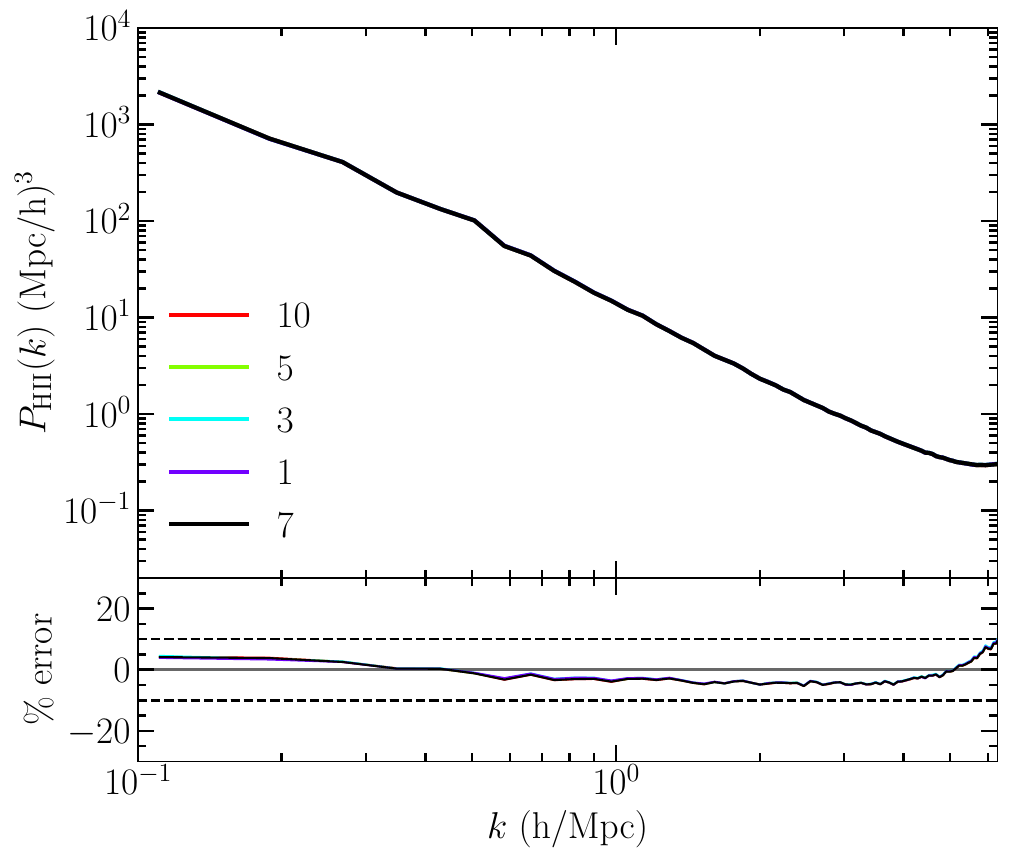}
    \label{conv_script_Q}    
  \end{subfigure}

  \vspace{-1em}
  
  \caption{\small HI-HI (left panel) and HII-HII (right panel) power spectra for different numbers of SB boxes combined for training, and the relative error of each with the true power spectra (lower panel), at fixed ionization fraction $Q^M_{\text{HII}}=0.5$. The default case that we work with is 7, shown in black. }
  \label{conv_script}
\end{figure}

\begin{figure}[h]
  \centering
  \begin{subfigure}{0.497\linewidth}
    \centering
    \includegraphics[width=\linewidth]{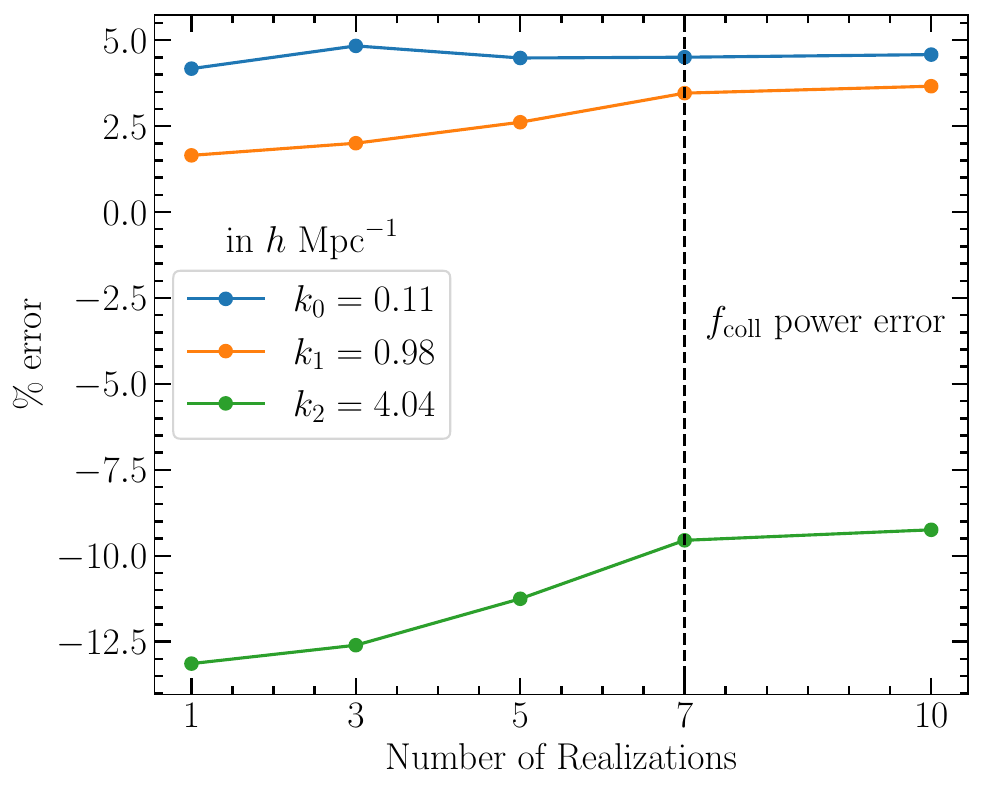}
    \label{conv_simnum_fcoll}
  \end{subfigure}
  \hspace{0em}
  \begin{subfigure}{0.483\linewidth}
    \includegraphics[width=\linewidth]{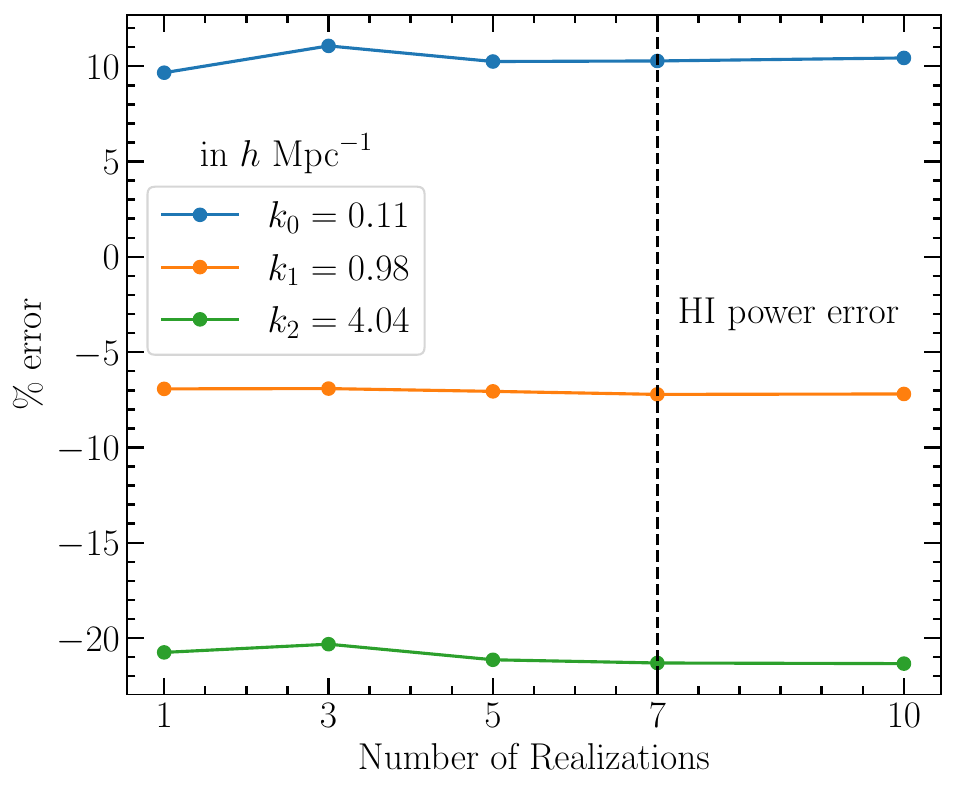}
    \label{conv_simnum_HI}    
  \end{subfigure}

  \vspace{-1em}
  
  \caption{\small Percentage error in the power spectra of \fcoll\ (\textit{left panel}) and HI density field (\textit{right panel}) against the number of SB realizations combined for training. The comparison has been made for three different $k$ values, corresponding to large $(k_0)$, intermediate $(k_1)$, and small $(k_2)$ scales. The black vertical dashed line highlights the number of realizations used in our method, and clearly demarcates convergence of the error to its right. The convergence trend is the most visible at small scales in the \fcoll\ power, where the error drops from $-13\%$ to $-9\%$. }
  \label{convergence_vs_sim_num}
\end{figure}

We chose to combine 7 different realizations of SB boxes to get the $(\delta, \fcoll)$ pairs, from which the training CDFs were constructed. Now, we vary this number to 1, 3, 5 and 10 and observe the effect on the results. The training converges successfully to $\verb|cv_thresh| = 0.015$ for each of these variations and the predicted \fcollm auto and cross power spectra are shown in Figure \ref{conv_fcoll}. In the lower panels, we show the error between the power spectra of each case with the truth obtained from RB, but the true power spectra itself is not shown in the upper panels. These results have been obtained for the fiducial $z=7$ setting. While the variation is small at large scales, a more clear trend can be noticed at the smallest scales, with the error curves of 7 and 10 realizations combined being almost identical. However, the \textsc{script} results are quite robust to these differences, and are similarly presented in Figure \ref{conv_script}. These convergence trends can be seen more explicitly in Figure \ref{convergence_vs_sim_num}, where we plot the variation of the power spectra error as a function of number of realizations, at three different $k$ values corresponding to large, intermediate and small scales.

This validates our choice of using 7 realizations to make the training, since any further increase in the number of realizations does not improve the results while any decrease causes the results to change, with the change being more significant for the \fcoll\ power. 


\section{Optimization of Binning}\label{appendix:optimization of binning}

As described in \ref{subsec:binning}, we adopt a variable binning scheme to construct the training data, defined over $\log(1+\delta)$. For the fiducial case, the bin widths for the first and last bins are 0.06 and 0.2 dex, respectively and a turning point occurs at $\delta= 0$, where it is the minimum at 0.03 dex. We first optimize the bin widths for the $z=9$ case, where in order to find the last bin width (high-density end), we start with a very small value which we apply to uniformly bin the whole $\log(1+\delta)$ range. We then train the GPR on the empirical CDFs constructed using such a binning. Then, we split the last bin into 3 equal sub-bins and evaluate the empirical CDFs for each sub-bin. These 3 CDFs are then compared with the 3 GPR-predicted CDFs at the central $\delta$ of these sub-bins. This procedure is repeated for successively larger parent bin widths. The idea is that at very small bin widths, GPR training at the last bin would suffer from noise and at very large bin widths the training CDFs would systematically deviate from the true underlying CDFs. In either case, the true empirical CDF at the 3 sub-bin centres would substantially differ from the GPR predictions. The parent bin-width that achieves visually the closest match is chosen as the optimal one, and it turns out to be around 0.12 dex for $z=9$. It must be noted that this was not a very strict choice, and slight deviations from this value do not appreciably change the results. The same procedure was carried out to find the optimal width of the first bin (low-density end) as 0.06 dex and the reference value of 0.03 dex at $\delta=0$. Given a $\delta$  binning, we find that 500 bins in \fcoll\ for making the training CDFs are usually enough to achieve accurate results for the joint distribution of $\delta$ and GPR predicted \fcoll. Still, we try another case with 900 bins just for comparison and choose the one with the better accuracy in recovering the HI power spectra, which for the $z=9$ case is the 900 bins case. Going beyond 900 does not improve the results.

Thus, once the $z=9$ binning is decided, we use it as a guideline to find the optimal bin-widths of the other cases. For $z=7$, we linearly scale the last bin width with $\log(1+\delta_{\text{max}})$, where $\delta_{\text{max}}$ is the highest density found in the SB at that redshift. This gives us a starting guess which we vary a few times along either direction. While $\delta_{\text{max}}$ changes substantially across redshifts, the minimum density values $\delta_{\text{min}}$ are very close to each other and hence we just try out a few variations along either direction of the $z=9$ first bin width (0.06 dex). These combinations are tested for both 500 and 900 bins in \fcoll\ for making the CDFs, and the case that gives the least error in the HI and HII power spectra is considered as the optimal choice.

\begin{figure}[h]
  \centering
  \begin{subfigure}{0.49\linewidth}
    \centering
    \includegraphics[width=\linewidth]{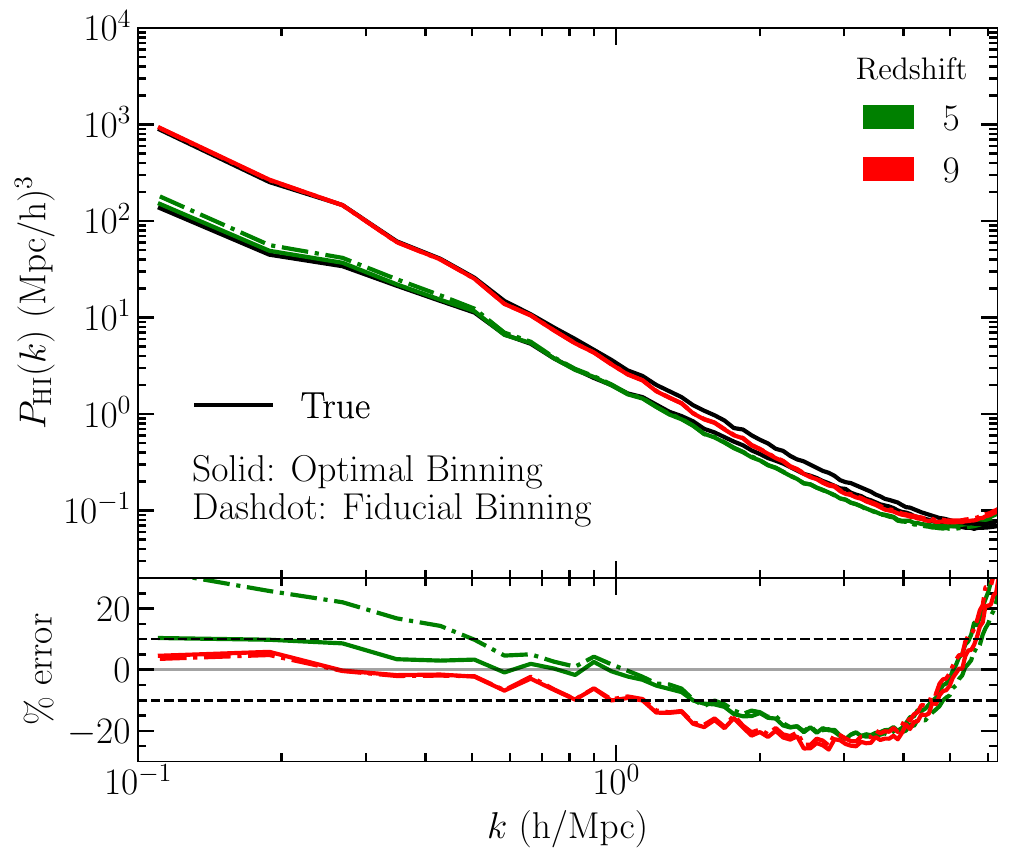}
    \label{z_var_HI_appendix}
  \end{subfigure}
  \hspace{0em}
  \begin{subfigure}{0.49\linewidth}
    \includegraphics[width=\linewidth]{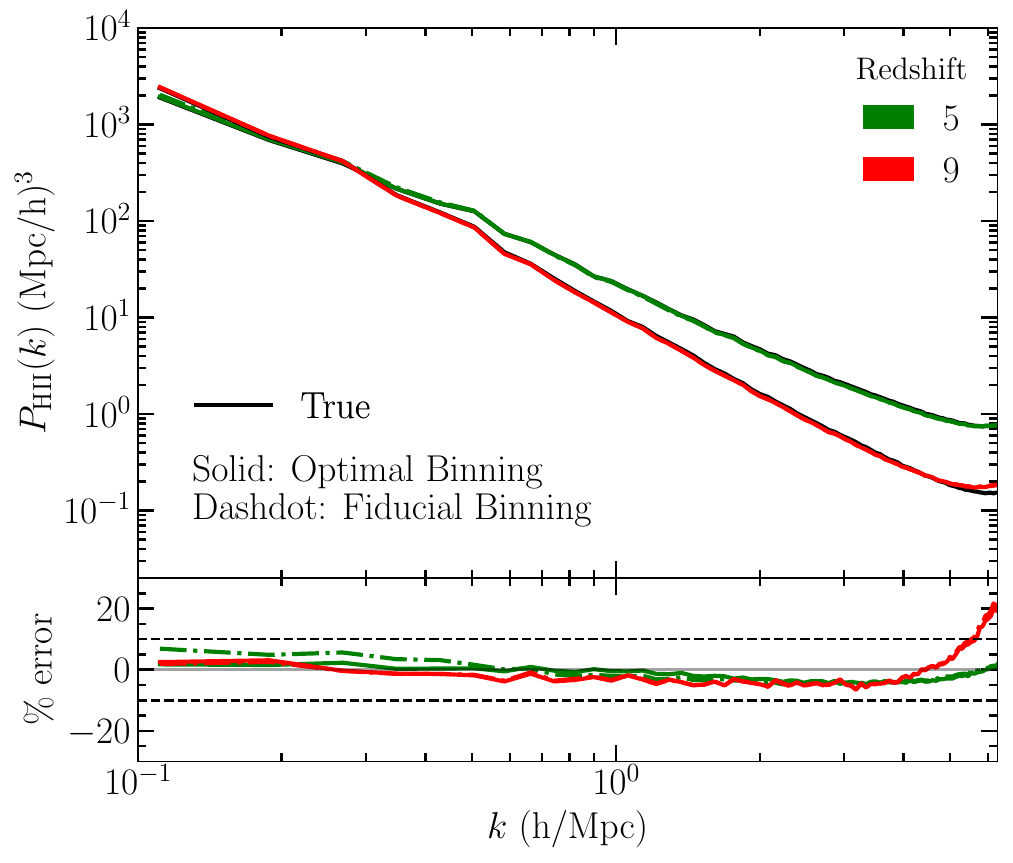}
    \label{z_var_Q_appendix}    
  \end{subfigure}

  \vspace{-1em}
  
  \caption{\small HI-HI (left panel), HII-HII (right panel) power spectra for truth and prediction using the interpolator made from (a) the fiducial (\textit{dashdot curve}) and (b) the optimized (\textit{solid curve}) binning schemes. The non-fiducial values of redshift are shown at fixed ionization fraction $Q^M_{\text{HII}}=0.5$. Only the $z=5$ case shows a significantly large error ($\sim 30\%$) compared to the case where its binning is separately optimized ($\sim 10\%$), and only at large scales of $k < 0.3\ \hMpc$ in the HI power. The difference at large scales in the HII power is not that stark, going from $\sim 1.5\%$ to $\sim 6\%$ over the same $k$ range. This highlights the importance of our optimization procedure. }
  \label{z_var_appendix}
\end{figure}

\begin{figure}[h]
  \centering
  \begin{subfigure}{0.49\linewidth}
    \centering
    \includegraphics[width=\linewidth]{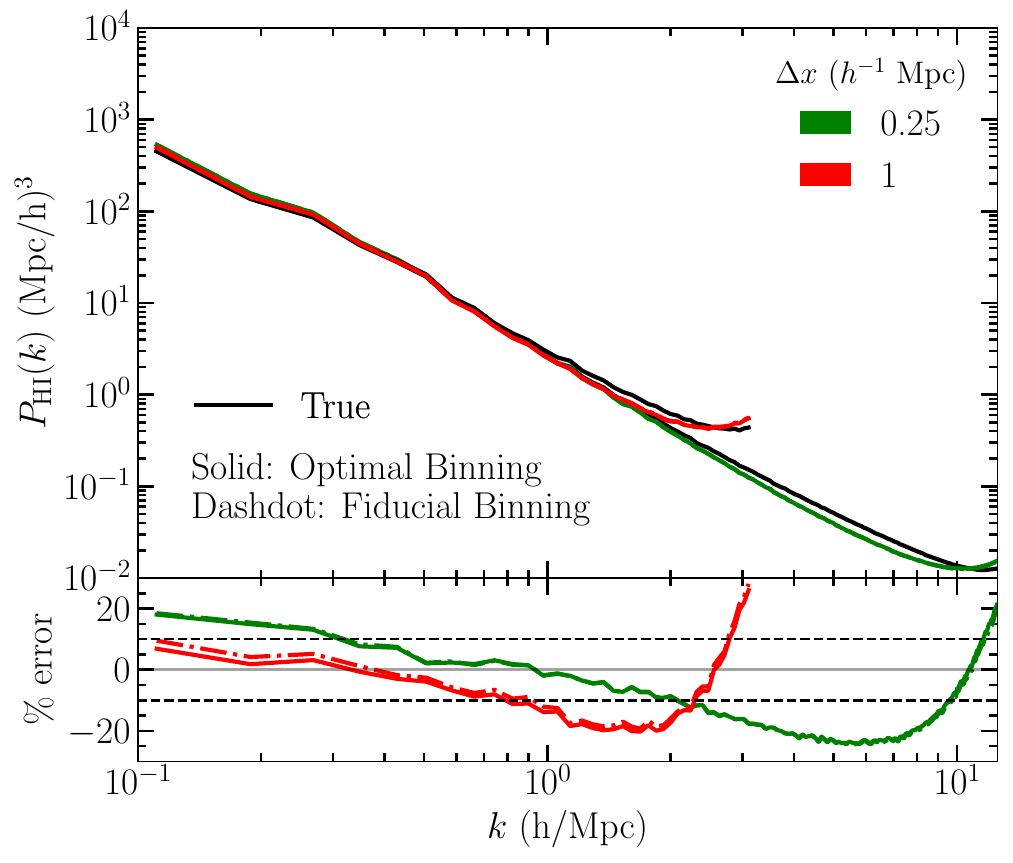}
    \label{deltax_var_HI_appendix}
  \end{subfigure}
  \hspace{0em}
  \begin{subfigure}{0.49\linewidth}
    \includegraphics[width=\linewidth]{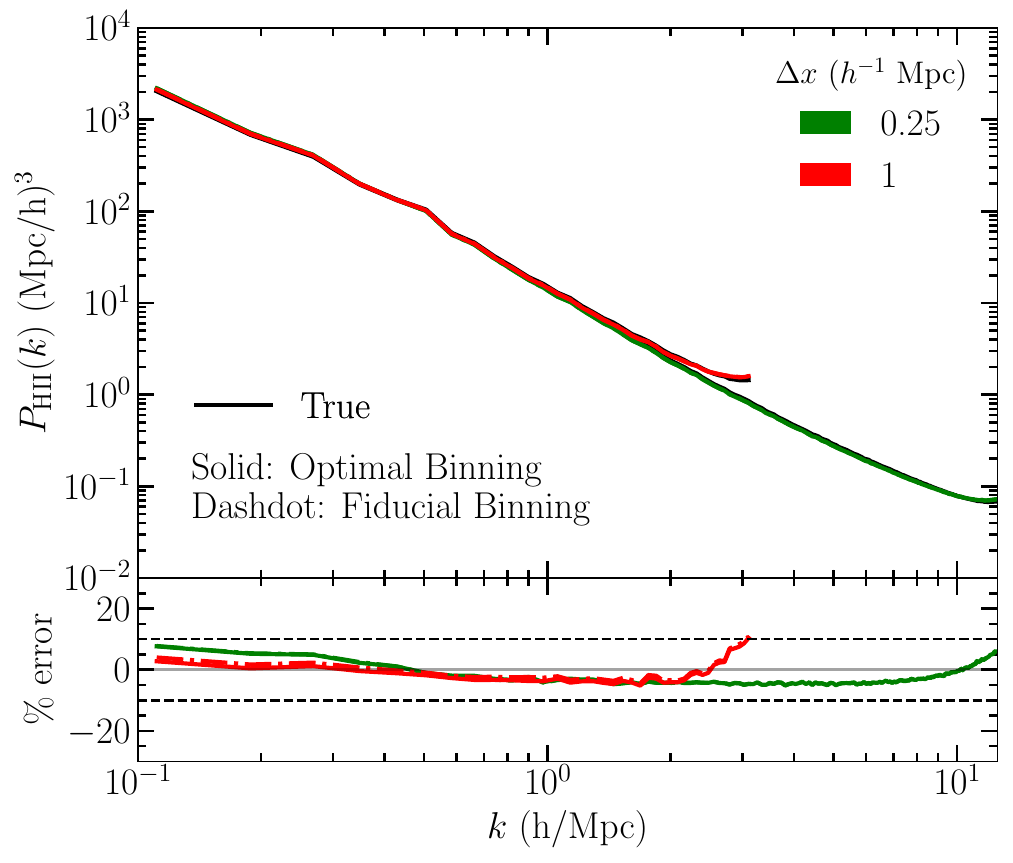}
    \label{deltax_var_Q_appendix}    
  \end{subfigure}

  \vspace{-1em}
  
  \caption{\small HI-HI (left panel) and HII-HII (right panel) power spectra for truth and prediction using the interpolator made from (a) the fiducial (\textit{dashdot curve}) and (b) the optimized (\textit{solid curve}) binning schemes. The non-fiducial values of grid size $\Delta x$ are shown at fixed ionization fraction $Q^M_{\text{HII}}=0.5$. Each case has been plotted upto its Nyquist frequency. The $\Delta x = 1 ~\Mpch$ case shows the most noticeable worsening of accuracy only in the large-scale HI power ($k < 0.3\ \hMpc$) with a $\sim 3\%$ degradation over its optimal binning scheme. }
  \label{deltax_var_appendix}
\end{figure}

If we just apply the binning scheme of our fiducial case on all the variations, the results worsen primarily for the redshift and the grid size variations. These are shown in Figures \ref{z_var_appendix} and \ref{deltax_var_appendix}, respectively, where we plot the HI power spectrum and its relative error with the truth using both --- the fiducial binning scheme (dashdot curve) and the separately optimized binning scheme for that case (solid curve). Evidently, in Figure \ref{z_var_appendix}, in the $z=5$ case, the HI power error worsens significantly, going from $\sim 10\%$ in the best interpolator constructed using the optimized binning to $>30\%$ in the interpolator with the fiducial binning, at the largest scales. The decline in the HII power accuracy is less severe for this case, going from $\sim 1.5\%$ to $\sim 6\%$. The $z=9$ case does not show any significant changes. For the gridsize variations in Figure \ref{deltax_var_appendix}, the $\Delta x = 1$ \Mpch case shows some noticeable deterioration going from $\sim 6.5\%$ to $\sim 9.5\%$ in the HI power error and $\sim 2.5\%$ to $\sim 4\%$ in the HII power error, at large scales. The change in the other case of $\Delta x = 0.25\ \Mpch$ is a lot less prominent. In general, we see that most of the results are not extremely sensitive to the binning scheme, especially at small and intermediate scales. The minimum halo mass variation cases have identical $\delta$ values as the fiducial case, and so applying the fiducial binning scheme over them does not result in any significant degradation, and are thus not shown here. 

\section{Normalization Error in \fcoll}\label{appendix:fcoll norm error}

\begin{figure}[h]
    \centering
    \includegraphics[width=0.49\linewidth, trim={0 0.7em 0 0.4em},clip]{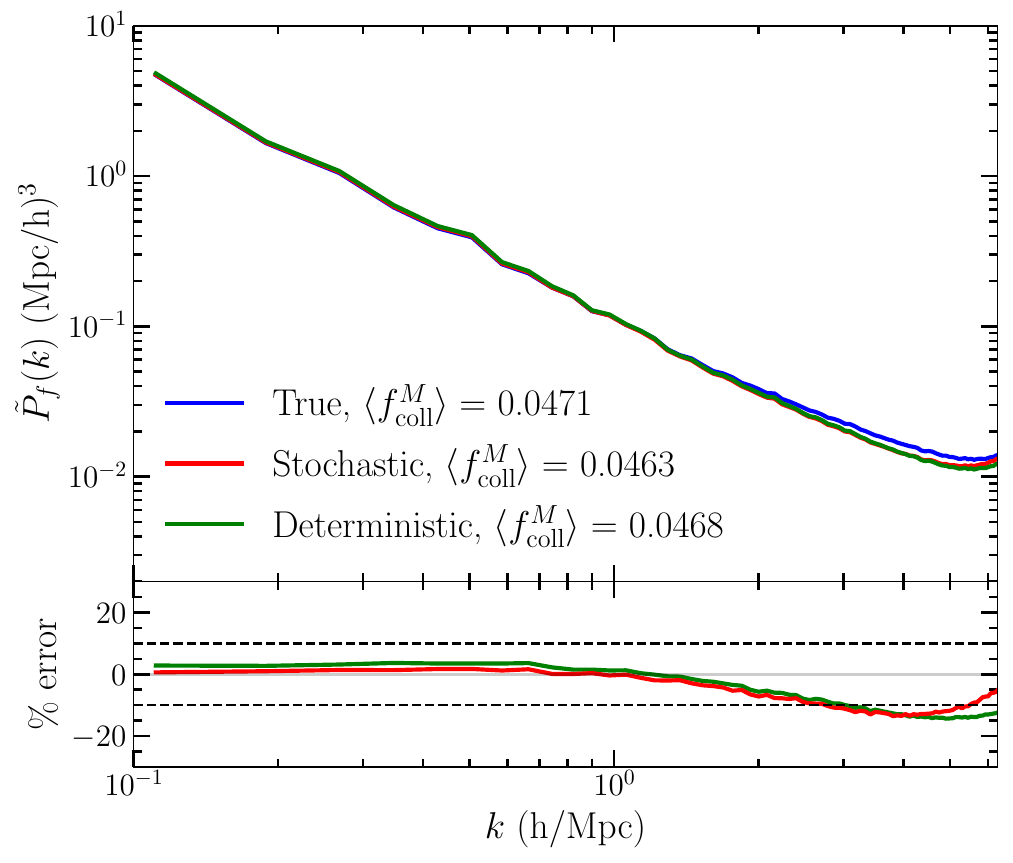}
    \caption{\small Comparison of $\fcollm$ \textit{unnormalized} auto power spectra (defined as per equation \ref{Pk_eqn_unnormed}), between truth and predictions using the deterministic and stochastic cases, for the fiducial parameter choice. The large-scale $(k < 0.7\ \hMpc)$ power matches quite well, with the deterministic error at $\sim 3\%$ and the stochastic error $\sim 1\%$, implying that the error in Figure \ref{fcoll_PS_z7} at large scales is mostly due to error in the mean.}
    \label{fcoll_PS_z7_unnormed}
\end{figure}

In sub-section \ref{subsec:fiducial}, we saw that the large-scale power of the normalized \fcollm\ field had a constant offset at around 5\% (Figure \ref{fcoll_PS_z7}), and mentioned that this can be traced back to the error in the global mean of \fcollm. Instead of using equation \ref{Pk_eqn}, we can define an unnormalized auto power spectrum (denoted by $\Tilde{P}_g(k)$) for a field $g(\mathbf{x})$ as -
\begin{equation}\label{Pk_eqn_unnormed}
\langle g(\mathbf{k})g^*(\mathbf{k}') \rangle = (2\pi)^3\Tilde{P}_g(k)\delta_D(\mathbf{k}-\mathbf{k}') \,,
\end{equation}
where the usual notations from equation \ref{Pk_eqn} apply. The difference is that we are no longer normalizing $g$ by its mean in position space while computing the Fourier conjugates. Comparing this $\Tilde{P}(k)$ for the stochastic and deterministic \fcollm in Figure \ref{fcoll_PS_z7_unnormed}, we see that both the predictions now show a very small error in the large-scale power. This shows that the 5\% offset at low $k$ in Figure \ref{fcoll_PS_z7}, at least for the stochastic case, is predominantly due to the error in recovering the global mean of \fcollm.

In the deterministic method, this error in the \fcollm mean comes from the effects of binning in $\delta$ and the fact that all training data comes from the SB, which intrinsically has a slightly different \fcollm mean than the RB. For the stochastic case, there is an additional error introduced by the GPR emulation of the CDFs. As seen in Figure \ref{joint_distri} the GPR-predicted global joint distribution of $\fcoll-\Delta$ does not exactly match the truth, and neither does the \fcollm mean.

\begin{figure}[h]
    \centering
    \includegraphics[width=0.49\linewidth, trim={0 0.7em 0 0.4em},clip]{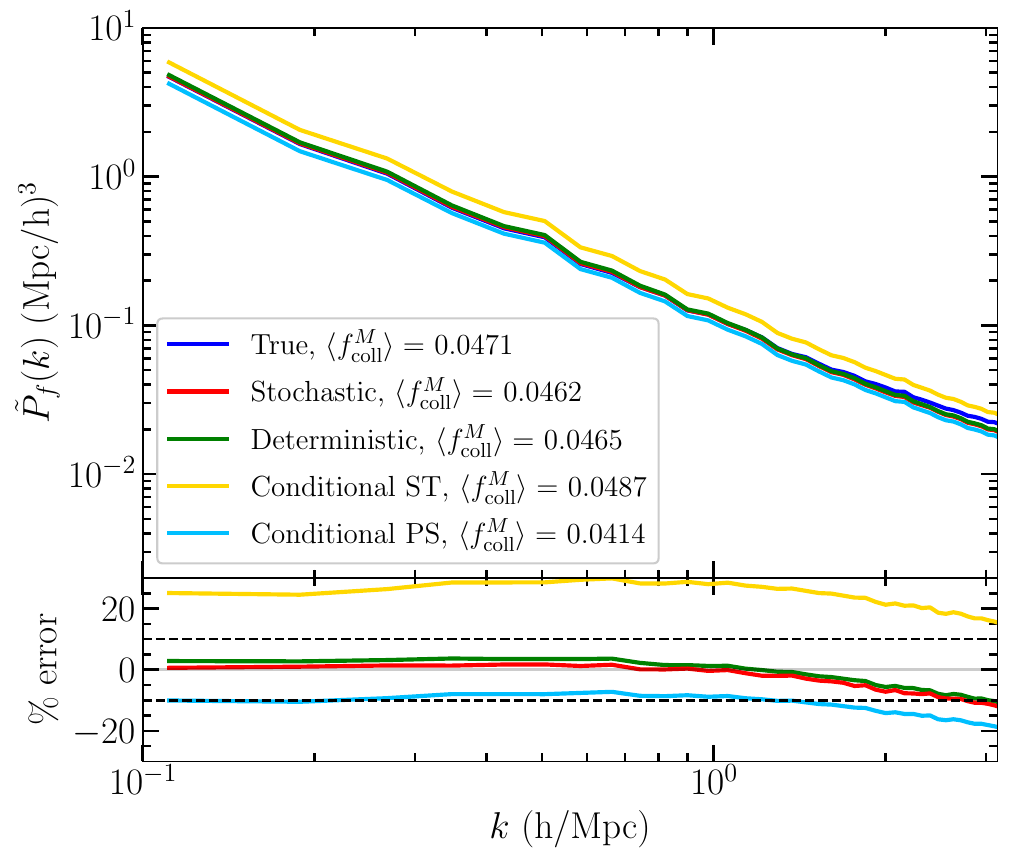}
    \caption{\small Comparison of $\fcollm$ \textit{unnormalized} auto power spectra (defined as per equation \ref{Pk_eqn_unnormed}), between truth, stochastic, deterministic and the semi-analytical predictions. Compared with the normalized case (Figure \ref{fcoll_PS_z7_full_comp}), the errors of conditional PS and ST are now very different (around $-10\%$ and $25\%$ for large scales of $k < 0.7\ \hMpc$, respectively) due their substantially different means.}
    \label{fcoll_PS_z7_full_comp_unnormed}
\end{figure}

We can also observe the effect of normalization on the power spectra of the conditional ST and PS predictions (Figure \ref{fcoll_PS_z7_full_comp}), by plotting their unnormalized $\Tilde{P}(k)$ in Figure \ref{fcoll_PS_z7_full_comp_unnormed}. As can be noted from the values mentioned in the plot, the conditional ST (conditional PS) overestimates (underestimates) the global \fcollm\ mean, with the difference being much greater than the stochastic and deterministic cases for conditional PS. Given the relation between the relative error in the normalized $(P)$ and unnormalized $(\Tilde{P})$ power spectra, 
\be
\dfrac{\Tilde{P}_{\text{sampled}}}{\Tilde{P}_{\text{truth}}} = \dfrac{P_{\text{sampled}}}{P_{\text{truth}}}\dfrac{\langle f_{\text{coll, sampled}}^M \rangle^2}{\langle f_{\text{coll, true}}^M \rangle^2}\,,
\ee
the unnormalized large-scale power has a larger magnitude of error in the case of conditional ST as compared to PS. When we move to the HI density field calculation, the $\zeta$ value required in order to achieve a global ionized fraction of $Q_{\text{HII}}^M = 0.5$ is then substantially different for the conditional PS and ST cases, which then ends up making a large error in their ionized bubble topology even at large-scales, hence causing the large $> 30\%$ error in large-scale power (left panel of Figure \ref{z7_script_full_comp}).

\end{document}